\documentclass{astlb}

\usepackage{graphicx}
\usepackage{natbib}
\usepackage{color}

\usepackage[hyperfootnotes=false]{hyperref}

\begin{document}

\journalinfo{2014}{40}{11}{667}[690]

\title{Measurements of the Cosmic X-ray Background of the Universe
and the MVN Experiment}

\author{M.~G.~Revnivtsev\email{revnivtsev@hea.iki.rssi.ru}\address{1}
  \addresstext{1}{Space Research Institute of RAS (IKI), Moscow, Russia}
}

\shortauthor{M.G. Revnivtsev}

\shorttitle{Measurements of the CXB and the MVN Experiment}

%\submitted{\today}
\submitted{May 22, 2014}

\begin{abstract}
The paper describes previous studies of the cosmic X-ray background (CXB) of the Universe in the energy range 1-100 keV and outline prospects for its investigation with the help of MVN (Monitor Vsego Neba) experiment. The nature of the CXB and its use for studying the cosmological evolution of black holes are briefly discussed. The bulk of the paper is devoted to the methods of CXB measurements, from the first pioneering rocket and balloon-borne experiments to the measurements made with latest-generation orbital X-ray observatories. Particular attention is given to the problems of allowance for the contribution of background events to the measurements with X-ray and hard X-ray instruments.
\end{abstract}

\section{1. INTRODUCTION}

The background emission of the sky in the X-ray
energy range (1-10 keV) is one of the first discoveries
in X-ray astronomy. The very first experiment whose
goal was the search for a non-solar X-ray emission
discovered two entirely new phenomena: the brightest source in the X-ray sky, Scorpius X-1 (a close
binary star system with a neutron star), and a nearly
isotropic sky emission, the cosmic X-ray background
of the Universe \citep{giacconi62}. At energies
below 1-2 keV, the background emission of the sky
is significantly anisotropic and has a different nature:
the hot-gas emission in our Galaxy (see, e.g., the
review by \citealt{mccammon90}.

The first experiments in X-ray astronomy were
carried out with high-altitude rockets capable of
bringing the recording equipment to altitudes higher
than 100 km (the experiments of the group from
the US Naval Research Laboratory, Friedman et al.,
are among the best-known ones; see the review by \cite{mandelshtam57}), on which the X-
ray absorption by the residual atmosphere is already
insignificant. The Sun was the object of observations
in the first experiments in the X-ray energy range.
The X-ray flux from the sky in these experiments
was observed with Geiger (gas) counters. The Sun’s
observations did not require a high-sensitivity of
instruments because of its considerable flux. Therefore, the X-ray flux from the Sun was generally
measured by integrating the total signal over certain
time intervals; no filtering of background events was
performed.

The main problem of recording the X-ray emission
of the sky is the background count rate that arises in
the detectors of this energy range even in the absence
of its real illumination by X-ray photons. Charged
cosmic-ray particles, mostly protons and electrons,
are the source of this background count rate.
Great progress in the sensitivity of gas counters to
(relatively) faint X-ray sources was achieved in 1962,
when an active anticoincidence shield was used to
separate the events due to the passage of charged
particles from the events related to the absorption of
X-ray photons; it allowed the background count rate
to be reduced by more than a factor of 100. This
led both to the discovery of discrete X-ray sources
in the sky and to the detection of a nearly isotropic
background emission.

The cosmic X-ray background was detected in the
original experiment of the ASE (American Science
and Engineering) group as a lower level of the ener-
getic particle count rate observed irrespective of the
direction of the instrument’s field of view. The authors
of the work provided arguments that these particles
were not charged ones but were X-ray photons \citep{giacconi62}.
The cosmic X-ray background (CXB) has since
been one of the main goals of the operation of any
orbital X-ray astrophysical observatory. A review
of the results of early experiments can be found in \cite{horstman75,tanaka77,boldt87}.

\section{2. THE COSMIC X-RAY BACKGROUND
OF THE UNIVERSE}

\begin{figure*}[t]
\begin{center}
\includegraphics[width=0.7\textwidth]{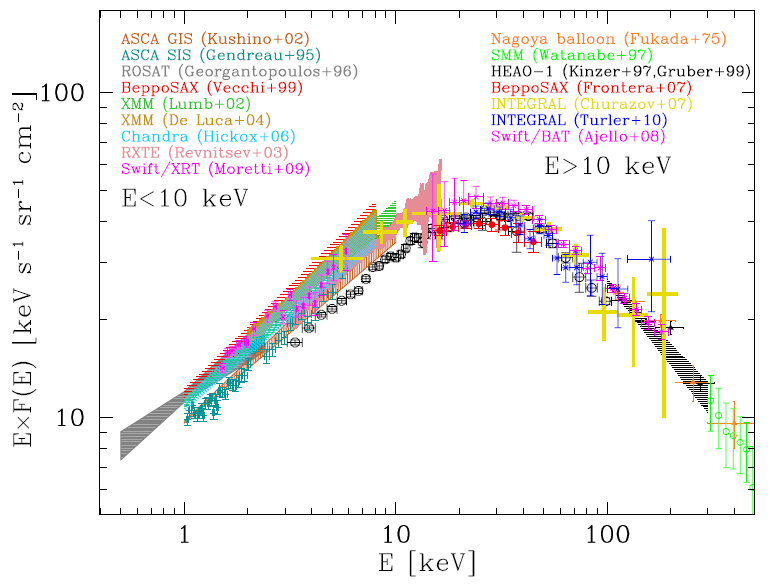}
\end{center}
\caption{\small Energy spectrum of the CXB of the Universe from the measurements of several orbital observatories. The maximum energy in the CXB is seen to be concentrated in the range 5-100 keV. The shape of the CXB spectrum can be satisfactorily described by a power law $dN/dE\propto E^{-\Gamma}$ with photon index $\Gamma\approx 1.4$ at energies below 10 keV and a photon index $\Gamma\approx 2.5$ at energies above 50 keV (from \citealt{gilli13}).}
\label{cxb_spectrum}
\end{figure*}

The first measurements of the CXB spectrum were
made by means of short (with an effective exposure
time of only  $\sim$300 s) rocket experiments in the energy range 1-6 keV and observations from stratospheric balloons at energies above 20 keV. In these
energy ranges, the CXB spectrum was individually
defined as a power-law dependence of the photon
number density on energy, $dN/dE\propto E^{-\Gamma}$. However,
it turned out that the measurements systematically
showed a difference in the slopes of the power-law
spectra at energies below and above 10 keV ($\Gamma\approx 1.4$
below 10 keV and $\Gamma\approx 2.5$ above 20-30 keV). Subsequently, higher-quality measurements of the CXB
spectrum showed that its hardness actually decreased
in the energy range 20-30 keV and, on the whole,
the CXB spectrum in the range 1-60 keV could be
described by the model of emission from an optically
thin plasma with a temperature of 40 keV. As a result
of these measurements, it was hypothesized that the
CXB was the result of emission from a hot intergalactic plasma with a temperature of 40 keV (see,
e.g., the reasoning in the early works of \citealt{felten66,silk70}).

Measurements of the shape of the cosmic microwave background (CMB) spectrum impose the
most stringent constraints on the existence of such
a hot intergalactic plasma. In the presence of a
hot plasma on the line of sight, the CMB photons
must undergo Compton scattering and gain energy
from energetic electrons in the intergalactic plasma.
As a result, the so-called $y$-distortion of the CMB
spectrum, the Sunyaev-Zeldovich effect, must be
produced (see, e.g., \citealt{sz70}).

The measurements made by the COBE orbital observatory showed the CMB spectrum to be an essentially Planckian spectrum of a perfect blackbody with the
possibilities of a deviation for the parameter $y$ by
no more than $y < 2.5 \times 10^{-5}$ \citep{mather94,wright94}. This means that the vast bulk
of the CXB flux cannot arise in the hot intergalactic
gas but must be produced by the total emission from
a large number of discrete sources.

The first direct measurements of a significant
contribution of the flux from a large number of
discrete sources to the CXB became possible after the advent of X-ray telescopes with grazing-incidence mirrors (first proposed by \citealt{giacconi60}). The first astrophysical X-ray telescope onboard the HEAO2/Einstein orbital observatory
(1978-1981) allowed a record (at that time) sensitivity of $1.3\times 10^{-14}$ erg s$^{-12}$ cm$^{-2}$ in the energy range 1-3 keV to be achieved. At this sensitivity level,
the surface density of detected sources was already
$\sim$20 sources deg$^{-2}$ \citep{giacconi79}.

At present, using the latest-generation telescopes
of the Chandra and XMM-Newton observatories, it
is possible to resolve more than 80-90\% of the CXB
flux at energies 1-2 keV into the contribution of
point sources with a surface density up to 10 000 per
square degree (see \citealt{moretti03,hickox06,hickox07}).

The overwhelming majority of the sources that
contribute to the observed CXB surface brightness
are active galactic nuclei (AGNs), accreting super-
massive black holes, at various distances from us \cite{setti89}. It is estimated that the
emission from ordinary galaxies and galaxy clusters
must contribute at energies below 2-5 keV. Since
the X-ray emission in AGNs arises from the accretion
of matter onto supermassive black holes, the CXB
surface brightness is actually an overall measure of
the growth of all supermassive black holes in the
history of the Universe \citep{soltan82,fabian99,gilli07,ueda14}.

Combining the CXB surface brightness measure-
ments with the studies of the counts of individual
classes of sources in various deep sky surveys allows
the long-term evolution of the growth of supermassive black holes to be studied.

A study of the latter showed that they could be
observationally separated into several large classes.
In the so-called unification model \citep{urry95}, the observational manifestations of AGNs in a
broad wavelength band (from the radio and infrared to
the X-ray range) depend on the AGN inclination to an
observer’s line of sight. The emission from the central
region of AGNs observed at an angle close to 90 ◦
passes through a dense dusty torus and is absorbed
up to the hard X-ray band, being reprocessed into
the infrared (the so-called type 2 AGN, Seyfert 2, absorbed AGNs). Manifestations of an accretion disk in
the ultraviolet and the soft X-ray band are clearly seen
in the spectra of AGNs observed at small angles to the
line of sight (the so-called type 1 AGNs, Seyfert 1).
AGNs observed along the direction of the relativistic
jet ejection from the central black hole are recorded as
“blazars” in a very wide wavelength range, from the
radio to ultrahigh-energy gamma rays (see, e.g., the
review by \citealt{boettcher07}).

Simulations of the properties of a population of
AGNs at various redshifts (see, e.g., \citealt{gilli07,sazonov08,ueda14}) show that
the AGNs surrounded by dust clouds with a column
density on the line of sight $\log N_H L > 23$ cm$^{-2}$ make
a large contribution to the CXB surface brightness at
energies above 5-7 keV (see Fig. 2).

\begin{figure}[htb]
\includegraphics[width=\columnwidth,bb=120 82 525 380,clip]{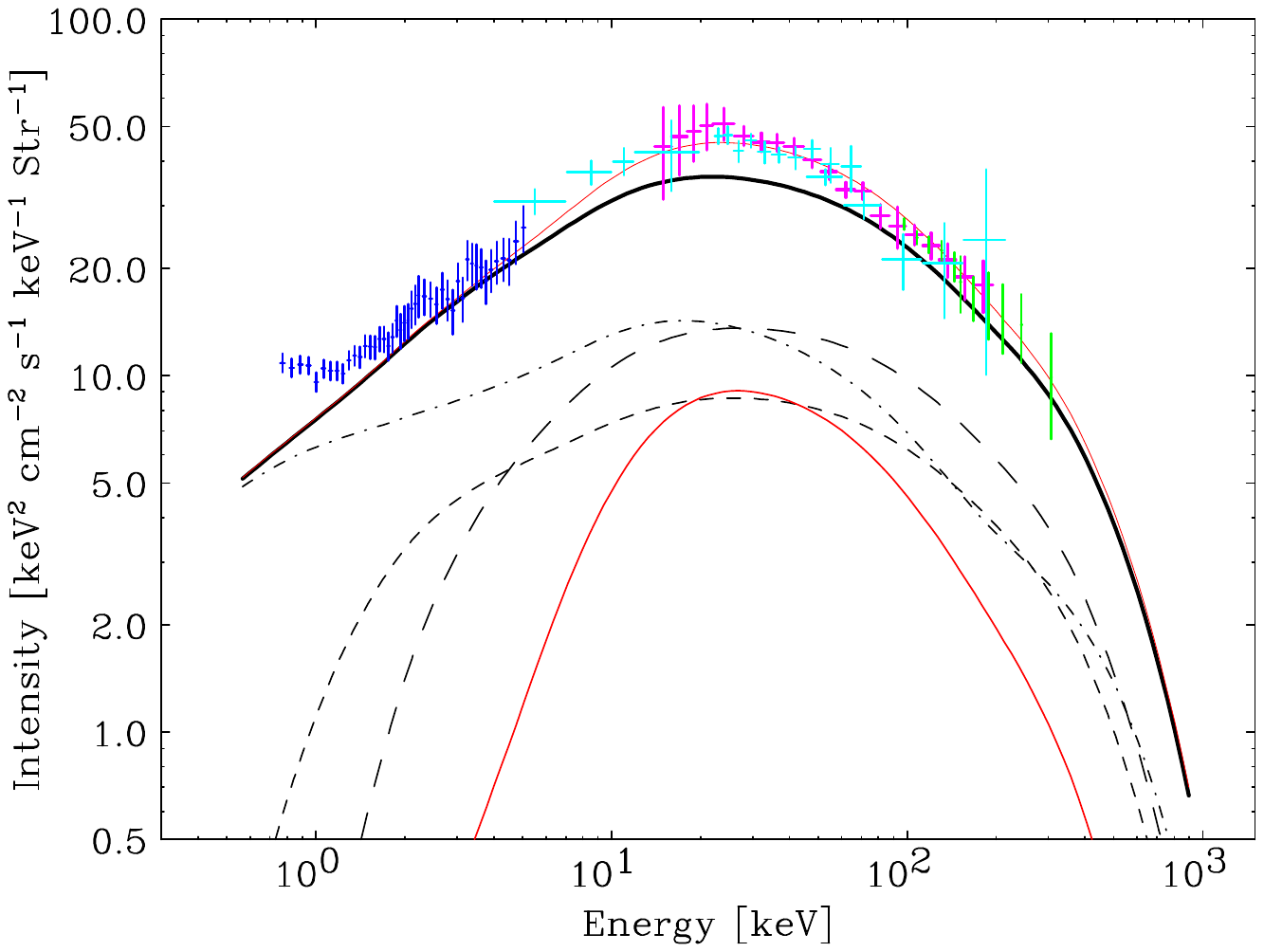}
\caption{\small Spectrum of the CXB of the Universe and its description by the contribution of AGNs with various degrees of absorption.
The upper thin solid curve indicates the complete model of emission from the sum of AGNs with absorption on the line of sight
from a column density $\log N_HL<22$ cm$^{-2}$ to $\log N_HL\sim26$ cm$^{-2}$. The lower thin solid curve indicates the contribution of
the so-called Compton-thick AGNs with a column density on the line of sight $\log N_HL=24-26$ cm$^{-2}$ (from \citealt{ueda14}).}
\label{cxb_population}
\end{figure}

The contribution of such AGNs and, hence, the
total density of black holes in the Universe can be
estimated accurately only by using information about
the shape of the CXB spectrum at energies above 5-7 keV.

An important feature of the CXB is that its surface
brightness contains the emission from all arbitrarily
faint objects in the Universe, even those that are
virtually impossible to observe separately with X-ray
telescopes because of their (telescopes) finite sensitivity.

This opens the possibility of measuring the total
(including the contribution of arbitrarily faint objects)
emissivity of the Universe. In the nearby Universe
(at distances up to 70-100 Mpc from us), such a
measurement can be made by using the property of
matter clumpiness on these spatial scales. Over the
evolution time of the Universe, the mutual attraction
of matter (including dark matter) gave rise to significant density inhomogeneities with a contrast of more
than unity. Since X-ray bright objects (AGNs, X-ray
binary systems, etc.) must trace the distribution of
baryonic matter, the X-ray volume emissivity of the
Universe must vary in space.

In observations of the X-ray sky, such variations in
the X-ray volume emissivity of the Universe must be
observed as increases and decreases of the CXB surface brightness in different directions: the distribution
of objects in the distant Universe will be averaged
over the directions, while the large inhomogeneities
in matter density at distances of 50-100 Mpc must
manifest themselves. If the distribution of matter (for
example, galaxies) is known, then the amplitude of
the CXB surface brightness variations can be recalculated to the total X-ray luminosity per unit volume
of the nearby Universe.

The first attempts at such measurements were
made using the sky survey with the A2 instrument onboard the HEAO1 observatory \citep{jahoda91,boldt92,miyaji94,scharf00}. The RXTE sky survey was used
for the same purpose in \cite{revnivtsev08}.
New measurements of the CXB surface brightness and its distribution over the sky with accuracies
better than 1-2\% are needed to accurately measure
the emissivity of the nearby Universe and to estimate
the contribution of the emission from faint sources
to it.

\section{3. THE PROBLEMS OF CXB
MEASUREMENTS}

The problem of CXB measurements can be separated into several components:

\begin{enumerate}

\item The problem of an accurate energy calibration of the instrument is related to the fact that
the CXB surface brightness decreases rapidly with
increasing photon energy. Therefore, even small errors in calibrating the energy scale of the detector will
lead to significant deviations of the measured CXB 
surface brightness from its true value. This question
is considered in more detail in Section 4.

\item The problem of allowance for the instrumental background of the detector is related to the
fact that the CXB is virtually isotropic (its fluctuations are no more than 7\% on a scale of one square degree and no more than 2-3\% on a scale of 20-40 deg.
(see, e.g., \citealt{schwartz70,schwartz76,warwick80,revnivtsev08}). Therefore, it
is very difficult to separate from the contribution of the
instrumental background of the detectors (for more
details, see Section 5).

\item The problem of an accurate absolute calibration of the measured flux is related to the limited
knowledge of the photon detection efficiency by the
detector being used and knowledge of its effective
collecting field of view. This problem is considered in
more detail in Section 6.

\end{enumerate}

\section{4. THE PROBLEM OF ENERGY
CALIBRATIONS}
\label{en_calib}

Inaccurate knowledge of the instrument’s energy
calibration inevitably leads to deviations of the measured CXB surface brightness from its true value.
For example, a 10\% error in the boundaries of the
energy channels for the CXB in the energy range 50-300 keV, where its spectral shape can be described by
a power law with a slope $\Gamma\sim 2.5$, will lead to an error
in the surface brightness of 25-27\%!

In the first experiments in the 1960-1970s, there
were only a few detector energy channels; there was
often no in-flight calibration, which is especially true
for the early experiments on the Soviet Kosmos satellites. Under such conditions, 10\% or larger uncertainties in the energy calibration were a common occurrence. For example, the attempts to recalibrate the
measurements of solar flares made by the scintillation
crystals on Prognoz satellites (5, 6, 7, 8) using the
measurements of the same events by the instruments
with an onboard energy calibration showed that the
declared energy scale of the detectors on the Prognoz
satellites could be shifted relative to the true one by a
factor up to 1.5-2! \citep{farnik84}

The possible variations of the instrument’s energy
scale under the action of various external factors are
an additional limitation of the accuracy of its energy
calibration. For example, one might expect a change
in the energy scale of a gas counter as the gas pressure in it, its temperature, or high voltage changes.

In the detectors based on crystal scintillators with
photomultiplier tubes, 10\% variations of the energy
scale due to the influence of the Earth’s magnetic
field were detected (the HEAO1/A4 experiment;
Jung 1989); on CsI(Tl) crystals with photodiodes and
based on CdTe semiconductor crystals, a temperature
dependence was detected (CsI(Tl), the PICsIT/IBIS
experiment of the INTEGRAL observatory \citep{malaguti03}; CdTe, the ISGRI/IBIS detector of the INTEGRAL observatory \citep{terrier03}.
The position-sensitive detectors based on CCD arrays have significant gradients in detector characteristics that should be measured during preflight calibrations and carefully traced during the in-orbit operation of the instrument (see, e.g., \citealt{dennerl02,plucinsky03,grant12}).

Thus, in order that the energy calibration of the
detector scale be maximally accurate, it is necessary
to provide maximally stable operational conditions
for the detectors and to systematically carry out its
on-board calibrations.

\section{5. THE INSTRUMENTAL BACKGROUND
OF DETECTORS}
\label{instrumental}

Measuring the isotropic emission of the sky is a
serious problem related to the separation of the flux
of X-ray photons arriving from the sky directly at the
detector from the count rate of all other particles.

The instrument recording X-ray photons actually
counts the following:

\begin{itemize}

\item  The nearly isotropic (independent of the direction of the instrument’s field of view) X-ray
background flux.
\item  The events caused by the passage of charged
particles (both Galactic cosmic rays and those
trapped by the Earth’s magnetic field) through
the detector.
\item The fluorescent X-ray photons produced in the
detector’s construction elements.
\item The X-rays and gamma-ray emission from the
Earth’s atmosphere. It can be both the reflected emission from the Sun/CXB/bright X-
ray sources and the emission produced by the
interaction of cosmic-ray particles with the atmosphere (the so-called Earth’s albedo emission; \citealt{vette62}). The Compton scattering
of gamma-ray photons in the detector body
can leave a slight energy in it, which will be
perceived as the recording of a low-energy X-
ray photon.
\item The X-ray and gamma-ray photons emerging
during the radioactive decay of construction
elements.
\item The X-rays (including those from the Sun)
scattered in construction elements.
\end{itemize}

Consider in more details the methods of separation and decrease all these background count rate components of the instrument.

\subsection{The Background Count Rate Components
of Instruments}

The instruments recording X-ray and hard X-ray
photons count the energy release events in the detector volume. The energy release can result from the
passage of both X-ray photons through the detector
body and various charged particles. For example,
on the Earth’s surface in the complete absence of
X-ray photons, the Geiger counters systematically
count about 0.5 count per second per cm$^{2}$ . The
passage of charged muons produced by the interaction of cosmic-ray protons (with energies above 1-10 GeV) with the Earth’s atmosphere constitutes the
overwhelming majority of the recorded events in this
case.

In the upper atmospheric layers, the muon flux
decreases, but the flux of Galactic cosmic-ray protons
and gamma-ray photons increases. The gamma-
ray photons result from the bremsstrahlung of relativistic cosmic-ray electrons and relativistic electrons
appearing during the decay of charged muons in the
atmosphere \citep{vette62,puskin70,petry05,sazonov08}.

Outside the atmosphere, the fluxes of charged particles are represented predominantly by high-energy
protons, electrons, and positrons.

\subsection{The Anticoincidence Shield}

\begin{figure*}
\hbox{
\includegraphics[height=0.9\columnwidth,bb=0 0 700 700,clip]{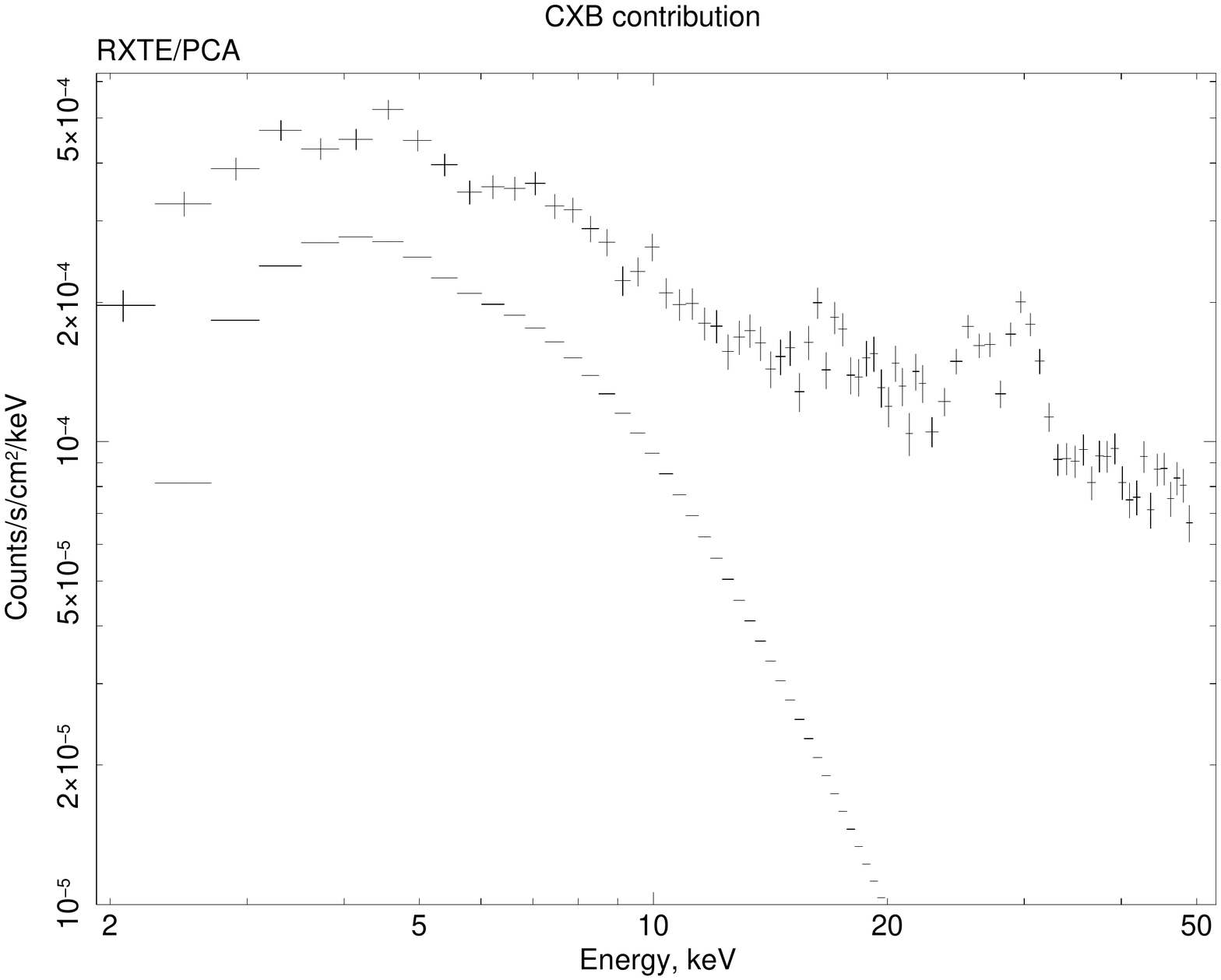}
\includegraphics[width=1.1\columnwidth]{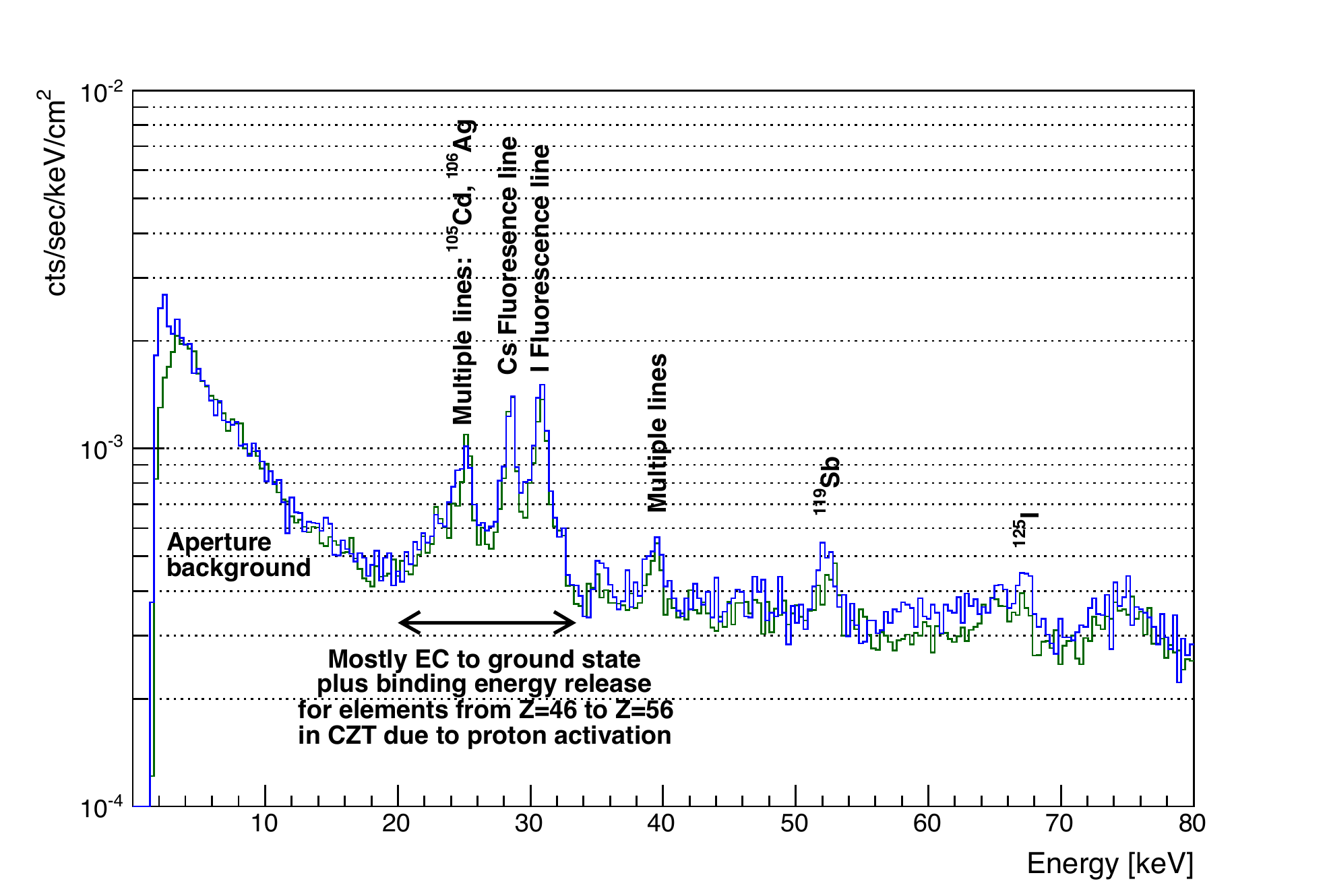}
}
\caption{\small Distribution of the event count rate in X-ray detectors in the energy left in the detector’s main body. In both cases, the
detectors were surrounded by an anticoincidence shield. (a) The spectrum of events in the PCA proportional counter of the
RXTE observatory. The histogram indicates the CXB contribution. (b) The spectrum of events when observing empty sky
fields by the CdZTe semiconductor detector of the NuSTAR observatory (from \citealt{harrison13}. The CXB contribution
dominates in the count rate at energies below 20 keV.}
\label{bkg_spectrum}
\end{figure*}

A passive shield of detectors with a surface density
up to 1 g cm$^{-2}$ efficiently stops the protons with
energies below 10 MeV.
The penetrating power of higher-energy protons is
large enough to pass right through the entire construction of the instrument. This property allowed
an efficient method of their filtering to be proposed.
The detector’s main body is surrounded by additional
(anticoincidence) detectors that record the passage of
high-energy particles.

The first versions of the anticoincidence shield
were based on plastic scintillators. X-ray photons in
the energy range 1-10 keV were completely absorbed
in the detector’s main body (for example, in the gas
counter), while high-energy charged particles also
passed through the scintillator. The simultaneous
triggering of the main detector and the active-shield
scintillator served as evidence that the occurred event
was related to the passage of a charged particle rather
than an X-ray photon.

Using multilayered gas counters was another way
of realizing an active anticoincidence shield. Several layers of charge-detecting anodes were passed
through the gas-counter volume. The layers of an-
odes located at the edges of the detector volume gave
a trigger simultaneously with the working layers of
the detector if an energetic charged particle passage
through the detector.

The efficiency of this method for the filtering of
background events in the energy range 1-30 keV
depends significantly on the type of the detector
used. In the case of a gas counter, the density of
the detector’s main body is fairly low; therefore, the
cosmic-ray protons with an energy of 10 GeV passing
through the detector leave an energy falling into its
operating range. For example, the gas density for a
xenon gas counter operating at a pressure of 1 atm
is $5.85\times 10^{-3}$ g cm$^{-3}$, and the mass of the material
traversed by the protons at a thickness of the active
gas layer 1-2 cm (as, for example, in HEAO1/A2,
EXOSAT/ME, GINGA/LAD, RXTE/PCA) is $\sim
7$ mg cm$^{-2}$. In this case, the energy lost by the
10-GeV cosmic-ray protons turns out to be $\sim 5-
15$ keV, falling within the operating range of the
instruments. Using an anticoincidence shield in the
case of gas counters allows the contribution of the
events unrelated to X-ray photons to be reduced by a
factor of 100.

The efficiency of the anticoincidence shield for
solid-state X-ray detectors (operating at energies be-
low 50-100 keV) decreases dramatically, because the
mean energy left by the high-energy protons in solid-
state detectors with densities of 2-8 g cm$^{-3}$ and
thicknesses of 1-2 mm (needed for the absorption of
photons with energies up to 70-100 keV) turns out
to be several hundred keV. Such events can be filtered
out just by the recorded energy release.

The high efficiency of the anticoincidence shield
for gas detectors led to the hasty conclusions that all
of the remaining events after its use were associated
with the real CXB. To illustrate that this is not the
case, Fig. 3 presents the spectrum of events recorded
in the PCA proportional counter of the RXTE observatory with an anticoincidence shield from all directions. The count rate of events unrelated to the
CXB is seen to be no more than 40-50\% of the total
one even at energies below 10 keV. Such a CXB
measurement error is inadmissible for further studies
of the cosmological evolution of black holes in the
Universe.

Thus, even when the anticoincidence shields are
used, the question about the separation of the contribution of events related to charged particles from the
count rate recorded by the detector remains a key one
for a reliable and accurate CXB measurement.

\subsection{Filtering by the Signal Rise Time}

Because of the great penetrating power of high-
energy charged particles, a charged particle in a geo-
metrically large detector ionizes a long trail that pro-
duces a long-duration signal in the detecting anodes,
considerably longer than that from X-ray photons.
This makes it possible to use information about the
rise time of the detector signal to separate the events
related to the passage of charged particles. This
method was proposed in the late 1960s \citep{gorenstein68} and showed a fairly high efficiency \cite{gorenstein69}.

For solid-state detectors with a large surface
density ($> 0.2-0.5$ g cm$^{-2}$), the efficiency of this
method is low for X-ray photons at energies below
100 keV. However, this method continues to be
used when working with hard X-ray and gamma-ray
photons at energies above 200-400 keV (see, e.g., \citealt{skelton00}).

\subsection{The Influence of Geomagnetic Cutoff Rigidity}

Not all of the events related to the passage of
charged particles can be filtered out by the anticoincidence shield even if it completely (in the solid angle $4\pi$) covers the recording instrument and through
filtering by the detector signal rise time.

The dependence of the measured count rate on
various orbital parameters, altitude, position above
the Earth, etc. is additionally studied to separate the
count rate components related to charged particles.
A major factor of the modulation of the count rate of
charged particles in this case is their reflection by the
Earth’s magnetic field.

Such a quantity as the magnetic rigidity $R$ can be
associated with any charged particle. Particles with
the same rigidity move along identical trajectories in
the Earth’s magnetic field:

$$
R={pc\over{Ze}}
$$

here $p$ is the momentum of the charged particle, $Ze$ is its charge, $c$ is the speed of light.

For a particle to be able to descent to a certain
altitude above the Earth, its magnetic rigidity must
exceed some threshold value. This value can be
calculated in the approximation of the Earth’s dipole
magnetic field:

$$
R_{\rm cut}={M \cos^4 \lambda\over{r^2 [\sqrt{1+\cos a \cos^3 \lambda}+1]^2}}
$$

here $\lambda$ is the geomagnetic latitude, $a$ is the entrance
angle of a positive particle ($\pi/2$ is the vertical motion), $r$ is the distance from the Earth’s center, and $M$ s the Earth’s magnetic dipole moment.

The geomagnetic cutoff rigidity can also be written
in the form of Stoermer’s simpler formula:

$$
R_{\rm cut}\sim 14.5 \times \left( 1+ {h\over{r_{\rm E}}}\right)^{-2} \cos^4 \lambda ~~\textrm{GeV}
$$

Here $h$ is the altitude above the Earth’s surface, and $r_{\rm E}$ is the mean radius of the Earth. The regions
near the equator have the largest geomagnetic cutoff
rigidity, $R_{\rm cut}\sim 20$ GeV.

If a particle has a rigidity smaller than that given
by the Earth’s magnetic field in this region, then
it cannot reach the Earth’s surface and will be reflected back into the space. Thus, it turns out that
the charged cosmic-ray particles with energies 1--20 GeV constituting the bulk of the cosmic rays in
interplanetary space reach the Earth’s surface differently; the fluxes of these charged particles are different
in different places of the Earth.
Both the rate of cosmic-ray passage through the
detector’s body and its illumination by the hard X-
ray and gamma-ray emission from the Earth’s atmosphere (arising from the interaction of cosmic rays
with the atmosphere) will depend on the magnetic
rigidity in which the instrument is at a given instant.

By investigating the dependence of the instrument’s count rate on magnetic rigidity, one can at-
tempt to separate the CXB contribution from the
charged-particle contribution.
The experiments on the Soviet Kosmos-135 (1966--1967), 163 (1967), and 461 (1971--1979) satellites \citep{golenetskii71,mazets75} can be
considered as an example of this approach. In these
experiments, an omnidirectional detector located at
a certain distance from the satellite’s main body (to
reduce the contribution from the induced radioactivity
of the satellite material) carried out measurements
in different parts of the orbit, at different magnetic
rigidities. The detector count rate was assumed to be
the sum of the CXB contribution independent of the
magnetic rigidity and the contribution of the Earth’s
induced emission dependent on the magnetic rigidity
of the region above which the satellite flies at a given
instant.

Subtracting the contribution of the count rate de-
pendent on geomagnetic cutoff rigidity allowed the
CXB to be estimated. In order to additionally get rid
of the contribution of the emission from the detector’s
induced radioactivity, the authors used the observations performed immediately after the satellite launch
until its first passage through the South Atlantic
Anomaly.

In fact, the described method allows only an upper
limit for the CXB surface brightness to be determined,
because the possibility that some part of the detector’s instrumental background is not modulated by
the geomagnetic cutoff rigidity remains anyway. This
may have become the reason why the CXB measurements from the Kosmos-135, 163, and 461 satellites
gave slightly larger surface brightnesses than other
experiments.

\subsection{The Absorption of X-ray Photons in the Upper
Atmospheric Layers}

In some rocket experiments, attempts were made
to separate the contribution of the events related
to the passage of charged particles from the events
related to X-ray photons by measuring their different dependence on the depth of the residual atmosphere. 

For example, in the experiments carried out
in September 1966 by a group from the Lawrence Radiation Laboratory of the US Department of Defence
and the University of California \citep{seward67},
it was pointed out that the count rate of the detector’s
anticoincidence shield was approximately the same at
altitudes of about 120 and 40 km. This was used as a
basis for the assumption that the effects related to the
passage of charged particles through the instrument
are identical at these altitudes (in fact, this is true
only partly, because the compositions of charged particles at these altitudes are fundamentally different;
therefore, their influence on the instrument can be
different). However, the flux of X-ray photons must be
absorbed by the residual atmosphere above the rocket
at an altitude of 40 km. Subtracting the instrument’s
count rates at these two altitudes gave the authors an
estimate of the CXB surface brightness.

\subsection{Radioactivity}

Energetic cosmic-ray particles are responsible not
only for the events recorded during their direct passage through the instrument. A very important component of the background event count rate for virtually any instruments is the count rate of particles resulting from the decay of radioactive elements
produced in the construction of the instrument or
satellite when they are irradiated by protons from the
Earth’s radiation belts. The concentration of such
protons is highest in the subpolar regions and in the
region of the so-called South Atlantic Anomaly, the
region in which the lower van Allen radiation belt is
closest to the Earth.

The specific set of elements passing into a radioactive state depends on the design of the spacecraft and
the detector.
After the passage through a region with a high
flux of energetic protons, the detector count rate rises
abruptly, although it must decline exponentially with
a characteristic time corresponding to the decay time
of the radioactive element (see, e.g., Fig. 4).
In some cases, the range of characteristic times
of the exponential decline in the instrument’s background count rate can be identified (as, for example,
for the LAC/GINGA \citep{hayashida89} and
PCA/RXTE \citep{jahoda06} proportional counters and NaI(Tl)-based scintillators \citep{jung89}.

In other cases, it turns out that the number of decaying radioactive elements is large enough and the
total background count rate declines already not exponentially but according to a power law \citep{dennis73}. In some experiments (for example,
OSO-3 and OSO-5), it was found that long-term
trends of an increase in the detector count rate occasionally exist, suggesting the accumulation of a large
number of long-lived radioactive isotopes.

\begin{figure}[htb]
\includegraphics[width=\columnwidth]{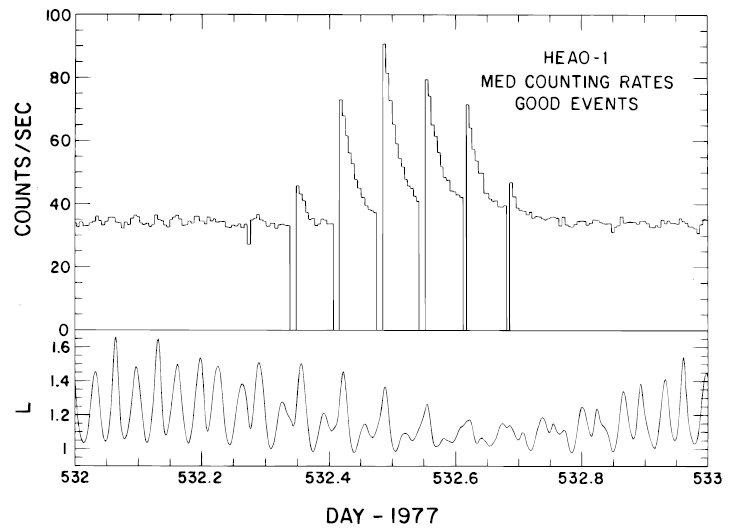}
\caption{\small Example of the count rate from the A4/MED detector (80-2000 keV) onboard the HEAO1 observatory on one of
the days. The background count rate is seen to increase sharply after the passage through the South Atlantic Anomaly. The
radioactive elements produced in this period (mainly $^{128}$I from the NaI scintillator crystal nuclei with a half-life of about 25 min)
then decay, leading to an approximately exponential decline in the count rate. From \cite{kinzer97}.}
\label{decay_kinzer97}
\end{figure}

The measurements made, for example, on OSO (Orbital Solar Observatory) satellites showed that
ignoring the contribution of the events due to radioactive elements (“activation”) led to a considerable
overestimation of the derived CXB surface brightness. An example of such measurements is shown
in Fig. 6.

\begin{figure}[htb]
\includegraphics[width=\columnwidth]{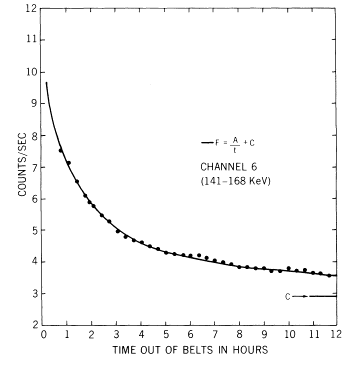}
\caption{\small Mean count rate of the detector in the energy range 141-168 keV as a function of the time since the last passage
through the South Atlantic Anomaly. The solid curve indicates a gradual decline in the count rate with time (from \citealt{dennis73}).}
\label{dennis_decay73}
\end{figure}

\begin{figure}[htb]
\includegraphics[width=\columnwidth]{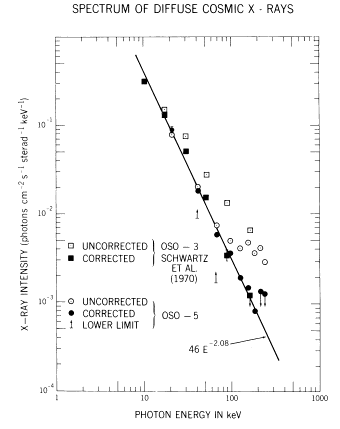}
\caption{\small CXB spectra from the results of OSO-3 and OSO-5 satellite measurements. The spectra of the events selected after
various filterings of the measurements are indicated by the open symbols. The filled symbols indicate the results of the CXB
measurements obtained after additional corrections for the possible contribution from the decay of radioactive elements (from \citealt{dennis73}).}
\label{sp_dennis73}
\end{figure}

\subsection{Blocking the Instrument’s Aperture}

The most obvious way to measure the contribution
of the events unrelated to the passage of X-ray photons from the sky through the instrument is a modulation of the instrument’s aperture, i.e., a periodic
blocking of the aperture by a passive or active shield
layer. If the aperture opening and closing cycle is
short enough, then it can be assumed that the back-
ground event count rate in these periods is the same,
but the instrument sees an additional contribution of
the flux of X-ray photons from the sky in the case of an
open aperture. A cover made of a material opaque to
X-ray photons in the energy range being investigated
can play the role of a passive shield.

One of the first such measurements was made in
late 1965 in a balloon-borne experiment. A NaI(Tl)-
based scintillator surrounded by an anticoincidence
shield scanned a ∼ 20 deg sky region, and its field of
view was periodically, with a period of 100 s, blocked
by a rotating 2-mm-thick tin cover. The CXB flux in
the energy range 26-90 keV was measured as a result
of these experiments (see Fig. 7).

\begin{figure}[htb]
\includegraphics[width=0.9\columnwidth]{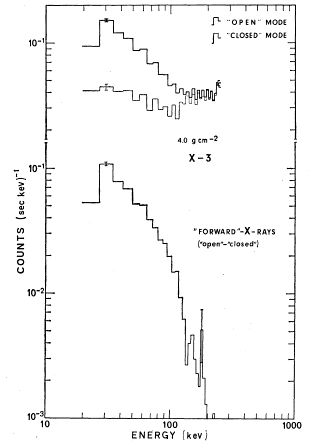}
\caption{\small Spectra of events measured with a scintillator when measuring the CXB in the regimes of an open and closed aperture
and their difference (from \citealt{bleeker70})}
\label{diff_bleeker70}
\end{figure}

Similar experiments on rockets were carried out
by a group from the Nagoya University \citep{fukada75}.
In late 1970-early 1971, a methodologically similar experiment was carried out onboard Lunokhod 1 during its operation on the lunar surface. The scientific equipment of Lunokhod 1 included the RT-1 collimating X-ray telescope consisting of two gas
counters with an effective area of 6.5 cm 2 (with a two-
layer system of anodes for the anticoincidence shield)
and a collimator limiting a 3.3-deg field of view. The
effective operating range of the detectors is 1-6 keV.
To separate the contribution of charged particles from
the total count rate of the detectors, the aperture of
the two counters should have alternately been covered
with a 10-μ iron filter blocking the X-ray photons.
Unfortunately, during the operation of the instrument
on the Moon, it turned out that the filter was not
transferred from one counter to the other, which did
not allow scientific results from this experiment to be
obtained.

Using identical detectors when observing the
same sky field but with one of them being blocked (for
example, by an additional absorbing layer) to prevent
the recording of X-ray photons can serve as another
realization of this approach. In this way, the CXB
surface brightness was measured on the Luna-12
interplanetary spacecraft in 1966 \citep{mandelshtam67}. In this experiment, the apertures of
two virtually identical Geiger counters were covered
with aluminium (10 $\mu$m) windows, but one counter
was additionally covered with a silver-gold foil filter
making it insensitive to X-ray photons. Comparison
of the count rates from the two counters made it
possible to estimate the CXB surface brightness in
the energy range 1-6 keV.

A significant shortcoming of this approach is the
limited identity of any devices. As a result, the actual
background count rates in two different detectors may
be considered identical only with a certain accuracy.
If this accuracy exceeds the amplitude of the useful
signal expected in the detector with an open aperture,
then this method of estimating the detector back-
ground turns out to be unreliable. For example, in
the Pulsar X-1 hard X-ray experiment on the Kvant
module of the Mir Space Station, an attempt was
made to use one of the NaI(Tl) scintillators to estimate the background count rate of other scintillators. The accuracy of this method turned out to be
unsatisfactory and, therefore, attempts were made to
determine the background count rate of the detectors
in real observations by shifting the field of view of the
detector with an open aperture, i.e., by alternating the
observations of the source and background fields (see,
e.g., \citealt{sunyaev94}).

\begin{figure}[htb]
\includegraphics[width=\columnwidth,bb= 1 477 393 771,clip]{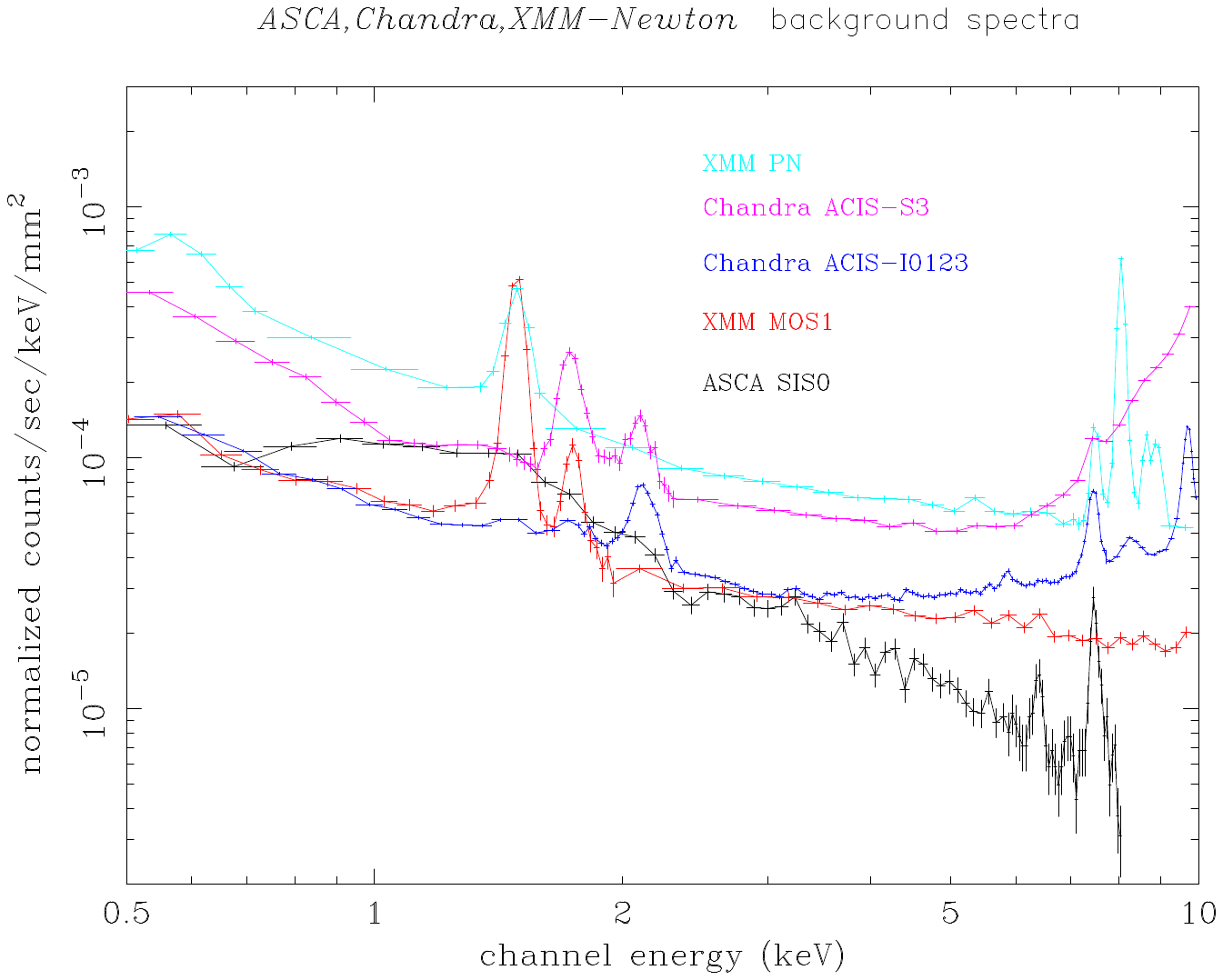}
\caption{\small Instrumental background spectrum for the ASCA, Chandra, and XMM-Newton detectors. The contribution from
the fluorescent lines of the detector materials is clearly seen at energies $\sim$1.5 (Al-K), 1.74 (Si-K), 2.123 (Au-M), 5.89, 6.49
(Mn-K), 7.48 (Ni-K), 8.05,8.91 (Cu-K), 8.64, 9.57 keV (Zn-K). The instruments are listed in the figure in order of decreasing
background intensity level at an energy of 5 keV (from \citealt{katayama04}).}
\label{fluorescence}
\end{figure}

\subsection{Blocking the Aperture by an Active Shield}

In the method with aperture blocking, there is
a possibility that a change in the configuration will
lead to a change in the count rate of the detector
background illumination. Theoretically, this is possible, because: first, high-energy cosmic rays create
induced radioactivity in the material of the detector or
satellite and in the material of the cover or collimators;
second, the cover or collimators can themselves be
the sources of X-ray emission, for example, through
fluorescence.

An example of the presence of fluorescent lines
from the material of the detector or its environment
can be seen in Fig. 8, which shows the actual instrumental background spectrum for the ASCA, Chandra, and XMM-Newton detectors (from \cite{katayama04}).

The fact that the appearance of an additional cover
in the detector aperture could affect the measurements at energies above several hundred keV was
demonstrated in laboratory experiments (for a description, see \citealt{kinzer97}). To reduce the influence of this effect, not only a passive shield blocking
the aperture but also an active shield was used in
some experiments. An element similar to the antico-
incidence shield, for example, CsI-based scintillators,
was used to cover the aperture. The events related to
the passage of charged particles through such a cover
are additionally cut off by the anticoincidence scheme.

One of the first such experiments was carried out
during the flight of a stratospheric balloon on Octo-
ber 17, 1973, by a group from the US Naval Research
Laboratory \citep{kinzer78}. Studies showed
that the effect from the activity (the anticoincidence
operation) of the cover blocking the detector aperture
at energies below 100-150 keV is small, less than
15\%, i.e., using an active or passive shield to block the
instrument’s aperture changed the derived CXB sur-
face brightness by no more than 15\%. More detailed
studies using the analogous А4/MED experiment
but installed on the HEAO1 satellite and, therefore,
making measurements for many months showed an
even smaller effect from the activity of the blocking
cover up to energies of several hundred keV \citep{kinzer97}.

\subsection{Modulation of the Instrument’s Field of View}

The instrument’s aperture can be changed not
only in variant 0 or 1 (open or closed) but also in
more steps. A two-step change in the instrument’s
field of view was made in balloon-borne experiments
by a group from the Nagoya University \citep{makino75}.
A four-step ($60^\circ\times 38.6^\circ$, $21^\circ\times 38.6^\circ$ ,$6.2^\circ\times 38.6^\circ$, and completely closed) change of the instrument’s field of view was made in a rocket experiment
in 1967 by a group from the Tokyo and Nagoya
Universities \citep{matsuoka69}. Measuring the
component whose flux depends linearly on the size of
the field of view open to the detector gave an estimate
of the CXB spectrum in the energy range 3.6-9 keV.
The field of view of the recording instrument was
modulated in more steps in a series of rocket experiments in the energy range 2-20 keV by a group from
the NASA Goddard Space Flight Center in 1968--1969 \citep{boldt69,boldt70}. In these experiments,
the gas counter was placed behind the collimating
system in which some (in the same direction) of the
collimating plates provided a $14^\circ$ field of view, while
the collimating plates in the perpendicular direction
could occupy 10 different positions, providing a set of
different effective solid angles, up to 0.17 steradians,
scanned by the detector in the sky. The flux directly
proportional to the solid angle in which the measure-
ments occurred gave an estimate of the isotropic CXB
surface brightness.

\begin{figure}[htb]
\includegraphics[width=\columnwidth,bb=25 249 560 530,clip]{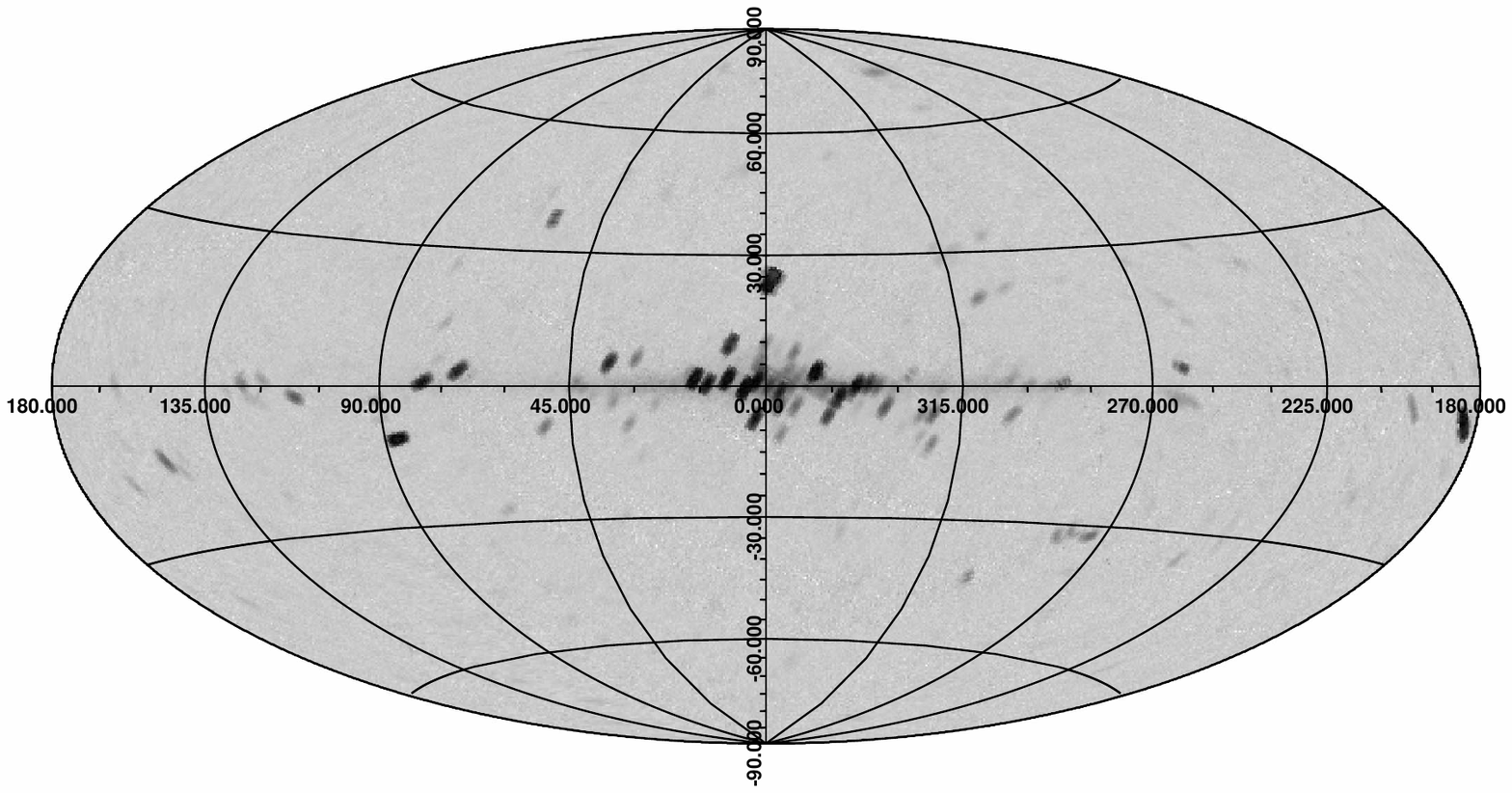}
\caption{\small 
Sky map from the measurements with the CXE/A2 instrument onboard the HEAO1 orbital observatory. The map is
shown in the Aitoff projection in Galactic coordinates. The straight line passing horizontally through the center of the map is
the Galactic equator. Since the angular resolution of the instrument is finite, the bright sources on the map are seen as $3^\circ\times6^\circ$
spots. The gray color in the bulk of the sky is the CXB accumulated by the CXE/A2 instrument in its $3^\circ\times6^\circ$ aperture.}
\label{skymap_a2}
\end{figure}

The technology of using a variable field of view
of the instrument to measure the CXB achieved its
greatest progress with the CXE/A2 experiment of the
Goddard Space Flight Center \citep{rothschild79}
onboard the HEAO1 observatory (1977-1979). The
instrument consisted of six gas counters that jointly
covered the energy range 0.15-60 keV. The entrance
aperture of the detectors was specified by a system
of collimators with an alternating size of their field
of view. The first and second halves of the charge-
detecting anodes were illuminated through the collimators limiting a field of view with a certain size
and a field of view approximately twice as large in
size, respectively. Two HED detectors (2.6--60 keV)
had $\sim 3 \times 3$ and $\sim 3 \times 6$ deg fields of view; two LED
detectors (0.15-3 keV), one MED detector (1.5-20 keV), and one HED detector had $\sim 3\times 3$ and $\sim 3 \times
1.5$ deg fields of view. Alternating the fields of view
above different anodes provided an almost identical
background count rate in these anodes at a CXB flux
differing by a factor of 2. Subtracting the measure-
ments of one set of anodes from the measurements
of the other set of anodes allowed one to make an
ideal subtraction of the instrumental background and
a reliable CXB measurement \citep{marshall80}.
These measurements are still deemed to be among
the most reliable ones in the energy range 2-60 keV.
The difficulty of determining the absolute calibration
of the measured flux remained a significant problem in
this case. Some time ago, an attempt was made to re-
calibrate these CXB measurements using the observations of the Crab Nebula \citep{revnivtsev05}.

\begin{figure}[htb]
\includegraphics[width=\columnwidth]{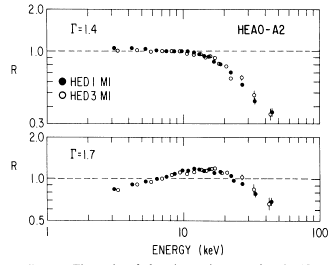}
\caption{\small Shape of the CXB spectrum from the measurements in the А2 experiment onboard the HEAO1 observatory. The ratio
of the spectral CXB surface brightness density to the model of its power-law energy dependence $dN/dE\propto E^{-\Gamma}$ is shown
(from \citealt{marshall80}).}
\label{cxb_marshall80}
\end{figure}

\subsection{Low-Energy Electrons and Protons}
The low-energy (below 1 MeV) electrons that,
when passing through the detector designed to record
X-ray photons, leave an energy falling into the instrument’s operating range are one of the important
components of the charged particle flux in space.
Thus, these electrons also produce the background
count rate in the instrument.

The problem of keV electrons was found back
in the early 1970s in rocket experiments (see, e.g., \citealt{seward74}).

The instruments designed to investigate the low-energy (below several keV) X-ray emission and,
therefore, having very thin aperture windows (for ex-
ample, formvar films with a thickness of 0.06 mg cm$^{-2}$)
were found to often record the fluxes of low-energy
electrons. The fluxes of such electrons influence
significantly the quality of cosmic X-ray emission
measurements (an example of this influence in the
rocket experiment on May 2, 1972, by a group from
the Livermore National Laboratory of the US Department of Energy is shown in Fig. 11). Therefore,
attempts were made to reduce their fluxes by various
methods, for example, by applying electrostatic and
magnetic fields to deflect the electron trajectories from
the detector.

\begin{figure}[htb]
\includegraphics[width=\columnwidth]{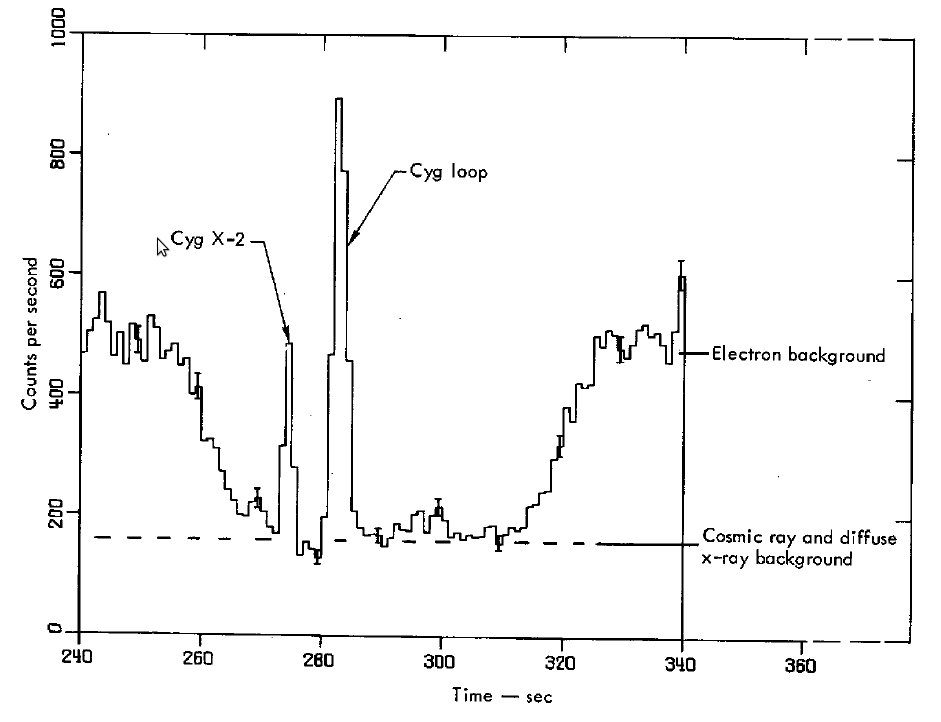}
\caption{\small Detector count rate in the rocket experiment on May 20, 1972, when scanning through the Cygnus region. The
background count rate produced by electrons is seen to be twice the background count rate from the X-ray background and
charged cosmic-ray. The detector aperture was covered with a formvar layer 0.035 mg cm$^{-2}$ in thickness. The detector count
rate in the energy range 0.2-4 keV is shown (from \citealt{seward74}).}
\label{eletrons_seward74}
\end{figure}

The numerous rocket experiments carried out under various conditions showed the following \citep{seward74}:

\begin{itemize}
\item  The electrostatic cutoff of electrons by applying
a voltage to the collimators does not give the
necessary effect. A high voltage applied to the
collimators leads to the generation of soft X-
ray photons, which creates more background
events than it allows them to be cut off.
\item The magnetic cutoff of electrons works very well. At magnetic fields up to 100 G, from
90 to 99\% of the electrons up to energies of
several tens of keV are cut off.
\item Highly polished collimators were found to additionally focus low-energy electrons into the
detector body. The rough collimator surface in
this case is a good way to combat this effect.
\end{itemize}

With the advent of telescopes with grazing-
incidence mirrors capable of focusing X-ray photons
with energies up to 10 keV, the problem of low-energy
charged particles became even more topical. Apart
from photons, such mirrors also efficiently focus low-
energy (up to several tens and hundreds of keV)
electrons and protons, increasing significantly the flux
of charged particles on the detector.

For example, such an increase in the flux of pro-
tons with energies of $\sim$100--200 keV led not only
to a rise in the detector background count rate but
also to a significant degradation of the receiving X-
ray CCD array mounted at the focus of the Chandra
telescope \citep{prigozhin00}.

Various configurations of magnetic deflectors are
currently used to reduce the flux of charged particles
with energies up to 100 keV. In particular, such deflectors were mounted on the optical system of the
ROSAT observatory; at present, they are mounted
on the optical systems of the Chandra, XMM, and
SWIFT/XRT observatories (for a description, see,
e.g., Spiga et al. 2008). The magnetic fields achievable at present on the small magnets of such deflectors (up to 1 kG) do not allow the problem of irradiation by protons with energies above 50-100 keV to
be solved.

Another way to solve the problem with the irradiation of a recording detector by low-energy
charged particles is to cover the entrance window
of the detector with a protective layer of material
capable of stopping the charged particles. In latest generation gas spectrometers, an additional propane
layer provides protection against low-energy electrons (see the RXTE/PCA detector scheme; \citealt{jahoda06}). The semiconductor detector based on
CdZnTe crystals onboard the NuSTAR observatory
is covered with a 110-$\mu$m-thick beryllium window,
which allows the electrons up to energies $\sim$100 keV
and the protons up to energies of less than 1 MeV to
be absorbed. On the ART-P X-ray telescope of the
GRANAT orbital observatory (1989-1998), a 500-$\mu$m-thick beryllium window performed the same role.

An interesting method for separating the contribution of low-energy electrons and photons with
energies 1-7 keV was applied in one of the detectors onboard the OSO-8 solar observatory (1975-1978). The entrance window of the gas counter of
detector B in the CXS (Cosmic X-ray Spectroscopy
experiment) experiment of the NASA Goddard Space
Flight Center was covered with protective layers of
different thicknesses. The entire detector was covered
with a 50-$\mu$m-thick beryllium layer. In addition, one
half of the detector was covered with a 30-$\mu$m-thick
beryllium layer, while the other half was covered with
a 20-$\mu$m-thick aluminium layer. These additional
thicknesses of the protective layers were chosen so
that the thickness of the entire protective layer in units
of g cm$^{-2}$ was the same. Thus, the protective layers
of the two halves of detector B in the CXS instrument
transmitted the low-energy electrons virtually identically and the photons with energies up to $\sim$7 keV
differently \citep{pravdo76}. The CXB was measured
using this difference.

\subsection{Blocking the Aperture by the Earth}
Using the Earth as a screen to cover the instrument’s aperture is an important way to separate the
count rate of events unrelated to the cosmic X-ray
emission and the CXB emission.
The Earth unlit by the Sun is a source of X-ray
emission due to the CXB reflection from its surface
and due to the emission arising from the interaction of
high-energy cosmic rays with the Earth’s atmosphere \citep{schwartz74,sazonov07,churazov07,churazov08,ajello08}.

The sunlit atmosphere of the Earth additionally
radiates due to the reflection of the solar flux (see
Fig. 12). It can be seen from Fig. 12 that the Earth’s
night side is a very faint source of X-ray emission up
to energies of several tens of keV. Thus, the Earth’s
night side can (1) serve as a very efficient screen
for the CXB, (2) give little intrinsic X-ray emission
(the surface brightness of the night-side Earth can
be lower than the CXB one by tens of times), and
(3) is also an opaque screen for cosmic rays, which is
very difficult to achieve with ordinary screens/covers
in instruments.

The observations of the Earth’s night side have
been used very actively to measure the instrumental
background of X-ray detectors since the early 1990s
(for example, for the ASCA \citep{gendreau95}
and BeppoSAX \citep{parmar99} orbital observatories; the Moon’s night side is used, for example,
to calibrate the background of the Chandra orbital
observatory \citep{markevitch03}.

In 2003, the observations of the Earth’s night side
were used to measure the CXB with the collimated
PCA spectrometer onboard the RXTE observatory \citep{revnivtsev03}.

In 2006, a special series of INTEGRAL observations of the Earth was performed to measure the
CXB in the energy range 5-200 keV \citep{churazov07}. The INTEGRAL observatory flies in a
highly elliptical orbit with an apogee of $\sim$150 000 km.
From such a distance, the Earth occupies only less
than 5 deg. in the sky, while the fields of view of
the main instruments are 15 and 20 deg. Therefore,
to maximize the useful signal (the total CXB flux
eclipsed by the Earth), the INTEGRAL observations
of the Earth were carried out in a period when the
distance from the Earth was 40 000--100 000 km \citep{churazov07}.

\begin{figure}[htb]
\includegraphics[width=\columnwidth,bb=85 1 700 580,clip]{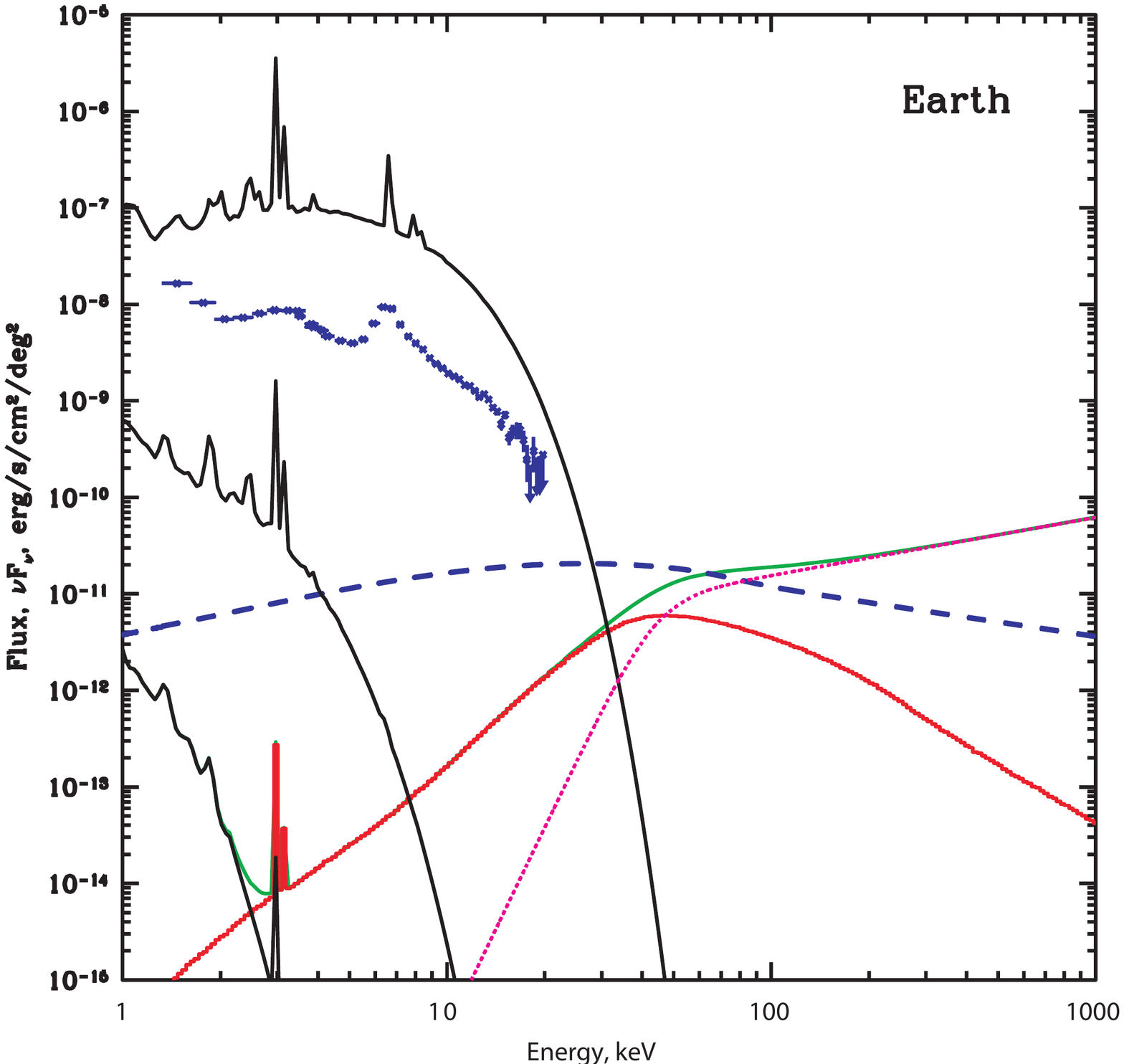}
\caption{\small X-ray spectrum of the Earth’s surface. The dashed curve indicates the spectrum of the incident CXB emission.
The lower solid curve indicates the contribution of the reflected CXB emission, the dotted curve indicates the contribution of
the so-called albedo emission arising from the interaction of high-energy cosmic rays with the atmosphere, the solid (black)
curves in the left part of the figure indicate the contribution of the solar emission reflected from the atmosphere at three levels
of solar activity: in quiescence (lower curve), at maximum solar activity (middle curve), and in a flare (upper curve; the possible
contribution from the nonthermal component of the solar emission is disregarded here). The asterisks indicate the spectrum of
the Earth’s day side in the case of an active Sun from the RXTE measurements (from \citealt{churazov07}).}
\label{earth_emission}
\end{figure}

At energies above 10 keV, the CXB measurements
using eclipses by the Earth are already complicated
significantly by the contribution of the Earth’s intrinsic hard X-ray emission (see Fig. 12). This contribution was taken into account by modeling the shape of
the Earth’s spectrum \cite{sazonov07,churazov07}.

A much stronger useful signal from the CXB
covering by the Earth can be measured on satellites in low near-Earth orbits (at altitudes of $\sim$500 km). Such measurements were made with the
currently largest hard X-ray BAT telescope onboard
the SWIFT observatory with an effective area of more
than 2000 sq. cm \citep{ajello08}. However,
despite the fairly high statistical significance of these
measurements, the technique itself used to determine
the CXB at energies above 10-20 keV suggests the
existence of a certain systematic error -- subtracting
the contribution of the Earth’s intrinsic emission is
required.

The conclusion follows from all of the aforesaid
that the CXB spectrum in the energy range 5--70 keV,
which is of great importance for determining the fraction of “hidden” (absorbed) AGNs in the Universe,
particularly at high redshifts, is currently known with
an unsatisfactory accuracy and requires a refinement.
It is especially important that the measurements
at energies below 10-15 keV and above 15-20 keV
are systematically carried out by different instruments
with different systematic errors and different absolute
calibrations. Therefore, properly joining the measurements in these energy ranges is a very important
problem when measuring the true CXB brightness
(for a discussion, see, e.g., \citealt{moretti09}).

\section{6. THE PROBLEMS OF ABSOLUTE
CALIBRATION OF CXB BRIGHTNESS
MEASUREMENTS}
\label{absolute_flux}

Once the contribution of the events unrelated to
the X-ray photons passing through the instrument’s
entrance aperture has been taken into account, it is
necessary to properly recalculate the recorded count
rate to the true CXB surface brightness.

There are several series problems in this way:

\begin{itemize}
\item The absolute calibration of the detector efficiency.
\item Determining the effective solid angle in the sky
from which the CXB flux is collected
\end{itemize}

\subsection{The Absolute Calibration of the Detector
Efficiency}

Determining the absolute detector efficiency is a
big technical problem for almost any X-ray instruments. As a result, the flux from the same astrophysical source measured at the same time turns out to be
different in different experiments.
To avoid this problem, it is necessary to carry out
a series of calibration measurements in laboratories
that would allow the absolute efficiency of the detectors, their correct effective area, etc. to be estimated.

The effective area of an instrument can often be
estimated with accuracies of 10-20\% through theoretical calculations of the photon detection process
by taking into account all the necessary physics and
geometry of the instrument. First of all, it is necessary to take into account maximally accurately the
distribution of events in the detector in energy channels dependent on the incident photon energy. Then,
it is necessary to determine the quantum efficiency
of the detector and its individual elements (if the
detector is a position-sensitive one). At present,
these calculations are often performed using software
packages specially developed to properly take into
account the various physical processes during the
interaction of photons (and other particles) in the
material, for example, the GEANT software package
http://geant4.cern.ch/ (for examples of calculations,
see \citealt{prigozhin03,godet09}).

Additional precise measurements are needed to
achieve an absolute accuracy better than 10\%. At
energies below 4-7 keV, such precise measurements
are often made using a synchrotron source of photons
whose absolute brightness can be accurately calculated given the current of electrons in the accelerator
beam used (see, e.g., \citealt{bautz00,krumrey04}). However, an absolute calibration accuracy for the instrument better than 5\% can rarely be
achieved even after these efforts.

At energies above 4-7 keV and for geometrically
large detectors, such measurements can no longer be
made. Instead, a set of different radioactive materials
with known values of their activity are used to mea-
sure the absolute efficiency of X-ray detectors (see,
e.g., \citealt{barthelmy05}).

Despite the great efforts put by the groups of
instrument developers to obtain good absolute calibrations of their instruments, it turns out that the
absolute accuracy of the measured fluxes from astrophysical sources is low in real experiments. One way
to combat this problem is to use some astrophysical
object with known characteristics to determine the
properties of an X-ray instrument. For example,
attempts are made to determine the parameters of soft
X-ray instruments in this way using the spectra of
hot white dwarfs and isolated neutron stars (see, e.g., \citealt{beuermann07}).

The cross-calibration of various X-ray instruments is often made by observing one of the brightest
sources in the X-ray sky, Tau Х-1 (the Crab Nebula).
This object is an isolated bright young pulsar in an X-
ray bright nebula (the Crab Nebula). The spectrum
of this source (pulsar+nebula) has no emission features and can be well described by a simple power
law with a photon index $\Gamma\approx2.1$ in a wide energy
range (1-100 keV). Numerous observations of this
source with various instruments show that its flux
is essentially constant and it can be effectively used
to test the calibrations of various instruments \citep{toor74,kirsch05}. It should be
noted, however, that small variations of its flux on a
scale of 4-6\% have recently been detected \citep{wilson11}, which should be kept in mind
when performing cross-calibrations with an accuracy
better than 5-7\%.

The source Tau X-1 gives a flux of $\sim 10^{-8}$ erg s$^{-1}$ cm${-2}$ and is too bright for almost
all latest-generation focusing X-ray telescopes. To
make the cross-calibration of these instruments,
attempts are made to use other astrophysical sources
whose flux should not vary significantly over tens
of years, for example, supernova remnants. Such
observations show that the problem of the calibration
accuracy for instruments still remains fairly serious:
for example, \cite{tsujimoto11} and \cite{ishida11} showed that the discrepancies in fluxes
from the same source could reach 20\% in the energy
range 2-8 keV and 46\%(!) in the range 15-50 keV.

\subsection{Determining the Effective Solid Angle}
An additional significant problem of determining
the CXB surface brightness is an accurate determination of the effective solid angle from which the CXB
photons are collected. Two main effects in this case
are vignetting and stray light.

Vignetting is a reduction in the flux from a source
recorded by the detector as it recedes from the center
of the field of view. In the case of detectors whose
field of view is limited by a system of collimators,
vignetting simply determines the instrument’s field of
view: the recorded flux from a point source in the
sky decreases linearly as it recedes from the center
of the field of view; the recorded flux from the source
at the edge of the field of view is zero. In the case
of focusing telescopes, the field of view is determined
by the detector size in the focal plane of the optical
system and by the properties of the focusing system.
Typically, the magnitude of the vignetting effect for
such systems is 70-80\% at the edge of the field of
 view. The vignetting effect is fairly easy to measure
by performing a set of observations for a constant
source in the X-ray sky at different positions in the
instrument’s field of view.

\begin{figure}[htb]
\begin{center}
\includegraphics[width=\columnwidth]{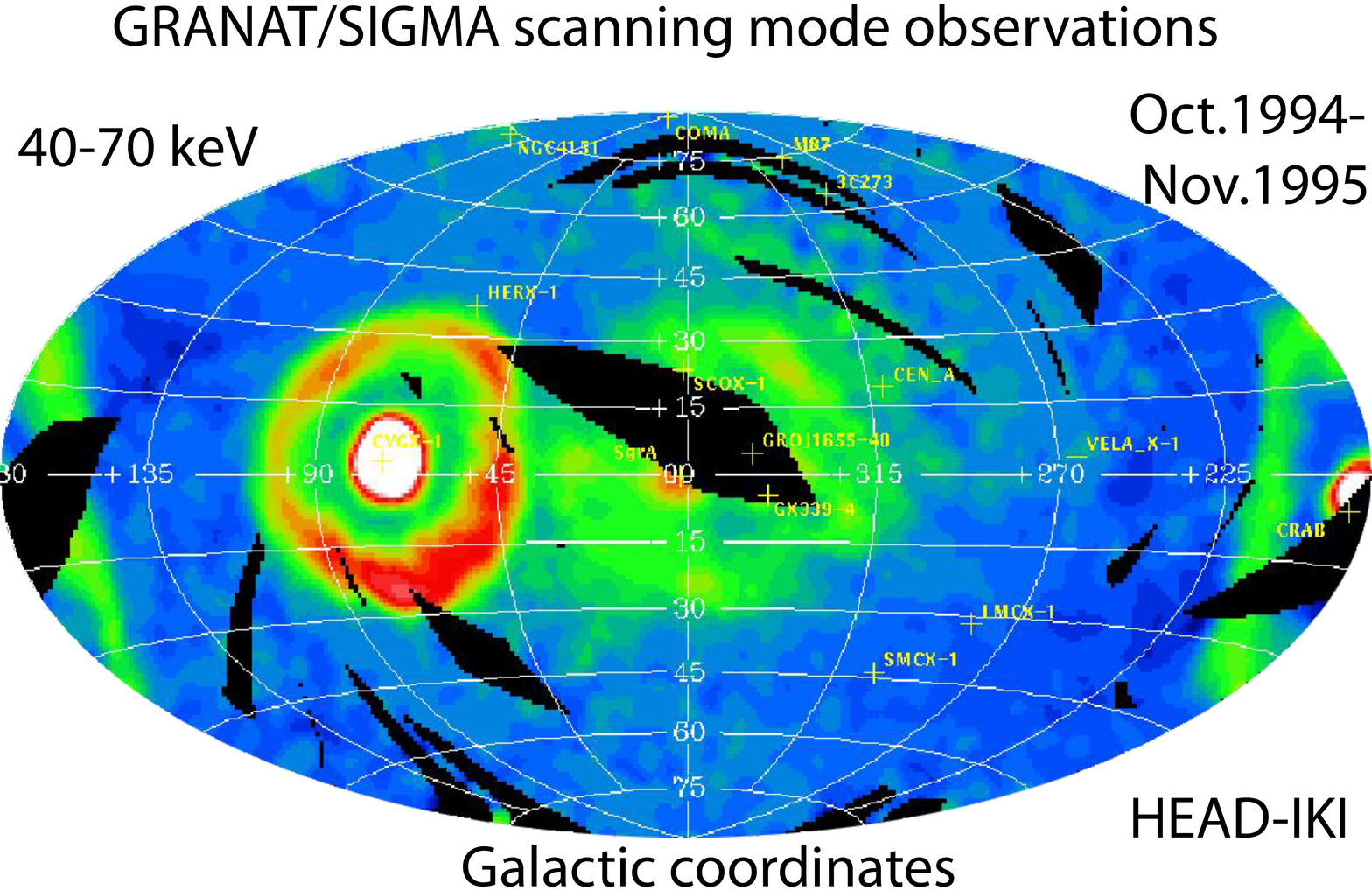}
\end{center}
\caption{\small Illustration of the existence of stray light in the hard X-ray SIGMA telescope. The all-sky map obtained with the
SIGMA telescope onboard the GRANAT observatory during scanning observations in 1994-1995 is shown (Churazov and
Gilfanov, private communication). Bright sources in the hard X-ray sky, the Crab Nebula and Cygnus X-1, are seen; the total
emission from the sources in the Galactic center region is seen. In addition, however, rings are clearly seen around the bright
sources arising from slits in the telescope’s shield.}
\label{sigma_scan}
\end{figure}

The stray-light effect is much more difficult to take
into account. It arises from the fact that the photons
arriving from regions outside the instrument’s field of
view and through the reflections that are not envisaged in its design fall on the X-ray detector.

For example, these can be the photons reflected
from the collimator walls (see, e.g., \citealt{dumas72})
or the photons that passed through the slits in
the detector’s active or passive shield. Figure 13
shows a good example of the presence of such stray-
light photons for the hard X-ray SIGMA telescope
onboard the GRANAT observatory. It can be seen
that apart from the photons incident on the detector
through the instrument’s main field of view (the spots
around the positions of bright sources in Fig. 13),
there exists a flux of photons incident on the detector
from a direction of $\sim$25-30 deg. from the center of
the main field of view (the circles around the positions
of bright sources). This flux is associated with the
existing slits in the telescope’s lateral shield (Claret
et al. 1994).

In focusing telescopes, stray light can arise due to
the direct (without any reflection from the mirrors)
falling of photons from the X-ray sky, due to the
photons that experienced one reflection (see, e.g., \citealt{chambure99}), or due to the photons
that arrived from outside the mirror system (as, for
example, in the NuSTAR telescope with an open
mirror system; \citealt{wik14}). Such stray light
manifests itself in the fact that the focal detector of
the X-ray telescope sees the X-ray photons from a
source far outside the telescope’s field of view. The
fraction of X-ray photons from sources outside the
field of view depends on its design, on the presence of
additional protective baffles. The stray light produced
on the detector of the ХММ-Newton telescope by a
constant source at various distances from the center
of the field of view is shown in Fig. 14.

It can be concluded that the stray light on the detector is produced by the convolution of the CXB surface brightness with the response of the
telescope+detector system dependent on the vignetting $f_{\rm vign}$, which reduces the effective solid angle
from which the CXB flux is collected, and on the stray
light $f_{\rm sl}$, which increases the effective solid angle from
which the CXB is collected, as follows:

\begin{figure}
\begin{center}
\includegraphics[width=\columnwidth,bb=55 109 423 381,clip]{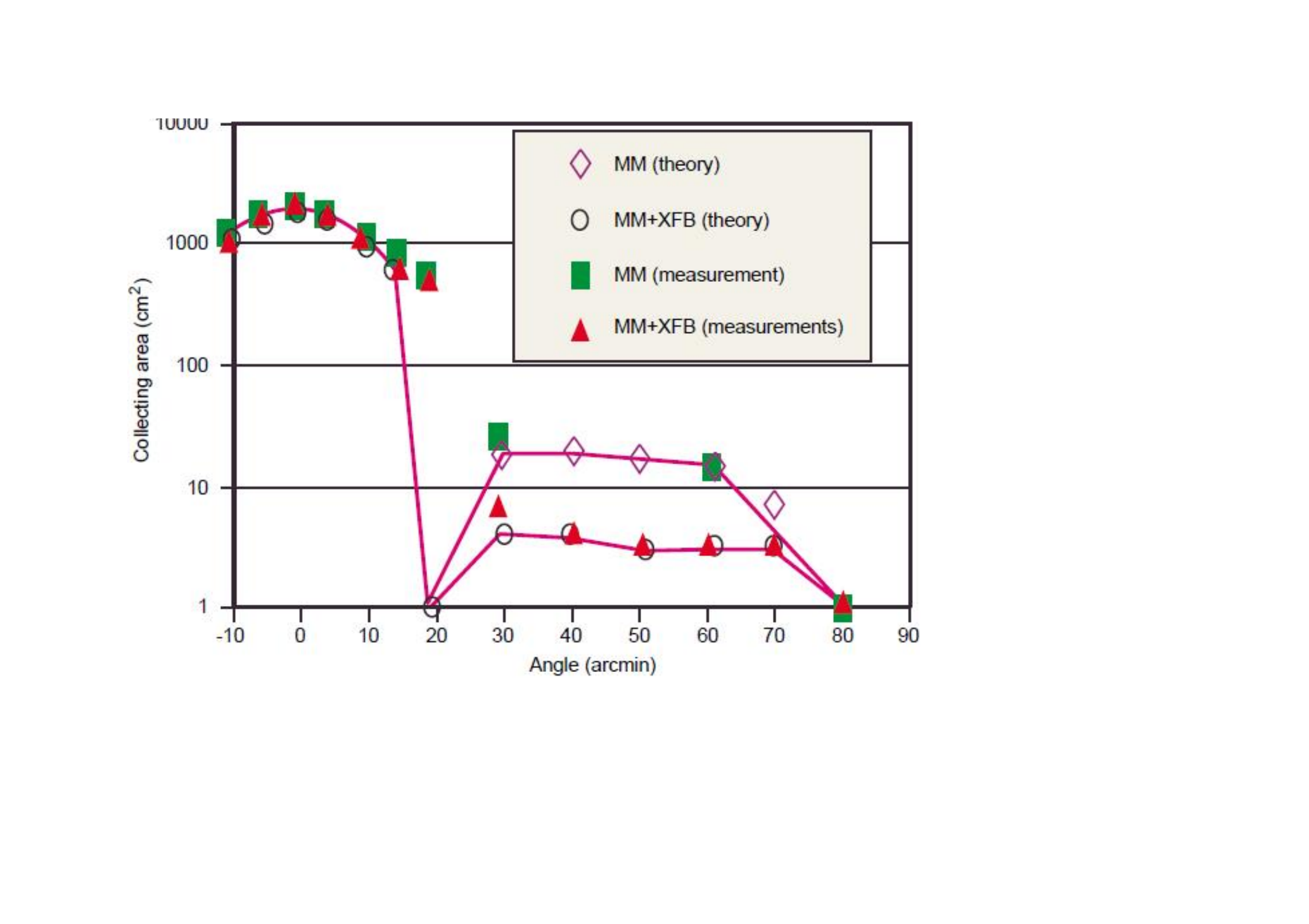}
\end{center}
\caption{\small Collecting area of one mirror system in the ХММ-Newton telescope. The mirror system is seen to continue to
concentrate the photons on the detector from sources far (up to a distance of 1.5 deg.) beyond the edge of the telescope’s field
of view (about 15 arcmin). From  \cite{chambure99}.}
\label{xmm_straylight}
\end{figure}

$$
F=\int{I_{\rm CXB}} ( f_{\rm vign} + f_{\rm sl}) d \Omega
$$

Since the CXB surface brightness is virtually in-
dependent of the direction in the sky, the convolutions
in this expression can be recalculated to a simple
product of the integrated vignetting and the integrated stray light with the CXB surface brightness.

Measuring these characteristics for a real instrument
is a very important factor determining the CXB measurement accuracy (see, e.g., \citealt{moretti09}).
Scanning observations of the sky often allow the
vignetting and stray-light effects to be measured di-
rectly from observational data by analyzing the fluxes
of constant sources in the sky, for example, the Crab
Nebula \citep{revnivtsev05,jahoda06}.
Special series of observations are performed for a
reliable measurement of these effects in mirror tele-
scopes, and the authors try to take them into ac-
count as accurately as possible (see, e.g., \citealt{moretti09}).

\section{7. THE MVN (MONITOR VSEGO NEBA) EXPERIMENT}

\begin{figure}[htb]
\begin{center}
\includegraphics[width=0.7\columnwidth]{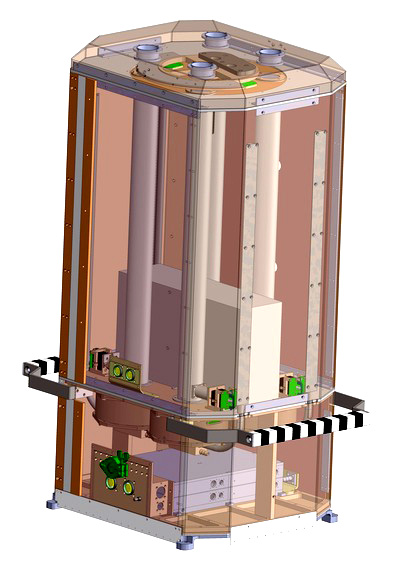}
\end{center}
\caption{\small General view of the MVN experiment. The main objective of the experiment is to measure the CXB surface
brightness with the maximum possible accuracy in the energy range 6-70 keV. The reliability of allowance for the instrumental
background of the detectors will be provided by periodically blocking the detector aperture by a rotating wheel (in the upper
part of the instrument); the stability of the energy scale and efficiency of the detectors will be controlled by the observations
of the calibration sources being periodically inserted into the field of view of the detectors (the inserts in the lower part of the
collimators). Thermal stabilization is created by the thermal control unit (the system control is located in the box between the
collimators).}
\label{mvn}
\end{figure}

The MVN (Monitor Vsego Neba, translates as the
all-sky monitor) experiment has been proposed by the
Space Research Institute of the Russian Academy
of Sciences and is being developed in collaboration
with the “Energia” Rocket and Space Corporation to
measure the CXB surface brightness in a wide energy
range, 5-70 keV \citep{revnivtsev12}.

The MVN instrument (see Fig. 15) is planned to be mounted
on the external platform of the Russian segment of the
International Space Station (ISS). Its orientation will
be fixed, and the instrument’s field of view will be directed to zenith. As a result of the evolution of the ISS
position with time, the instrument’s field of view will
gradually cover different areas in the sky and and will
cover the entire reachable part of the celestial sphere,
$\sim$83\% of the entire sky or $32.4\times 10^3$ sq. deg., over
a period of about 72 days. The reachable part of the
sky will be covered $\sim$15 times over three years of the
instrument’s planned operation.

\begin{figure}[htb]
\begin{center}
\includegraphics[width=\columnwidth]{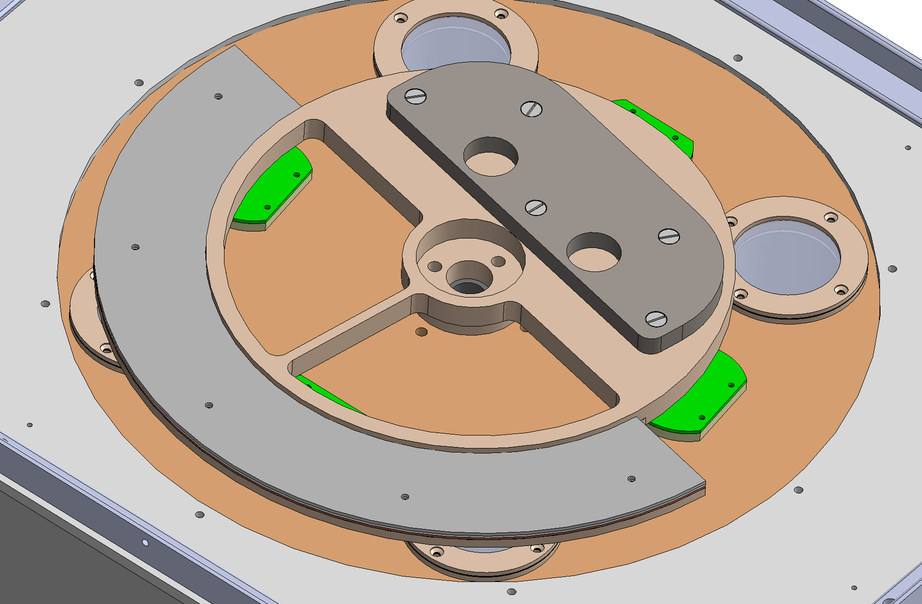}
\end{center}
\caption{\small View of the entrance windows of the collimators in the MVN experiment and the rotating modulating wheel.}
\label{mvn_koleso}
\end{figure}

The main parameters of the MVN experiment are:

\begin{table}[htb]
\caption{Main parameters of MVN experiment} 
\label{pars}
\begin{center}       
\begin{tabular}{ll} %% this creates two columns
%% |l|l| to left justify each column entry
%% |c|c| to center each column entry
%% use of \rule[]{}{} below opens up each row
\rule[-1ex]{0pt}{3.5ex}  Crystal & CdTe  \\
\rule[-1ex]{0pt}{3.5ex}  Crystal thickness & 1 mm  \\
\rule[-1ex]{0pt}{3.5ex}  Energy band & 5-70 keV  \\
\rule[-1ex]{0pt}{3.5ex}  Field of view & $\sim$3.2$^\circ$ diameter \\
\rule[-1ex]{0pt}{3.5ex}  Effective area & $4\times 4.5$ cm$^2$ \\
\rule[-1ex]{0pt}{3.5ex}  Operational temperature & $-30^\circ$C \\
\rule[-1ex]{0pt}{3.5ex}  Time resolution & 1 мсек \\
\rule[-1ex]{0pt}{3.5ex}  Power consumption & 40 W \\
\end{tabular}
\end{center}
\end{table} 

The MVN instrument consists of four identical
semiconductor CdTe detectors placed under cylindrical collimators limiting a field of view 3.2 deg. in diameter. The collimators are made of three metal layers, each with a thickness of 1 mm; from the outside
inward, the layers are tin, copper, and aluminium. The
sequence of metals in the collimators was chosen so
that the fluorescent lines arising in the outer parts of
the collimator were absorbed in the inner ones, while
the fluorescent lines of the inner layer (aluminium at
an energy of $\sim$1.56 keV) were outside the operating
range of the instrument. The detector crystals are
surrounded by the same multilayered passive shield
with a surface density of $\sim$1.9 g cm$^{-2}$. The shield
transparency is no more than 3\% for photons with
energies below 70 keV.

Since the goal of MVN is to measure the cosmic
X-ray background maximally reliably and accurately,
all the difficulties of CXB measurements described in
the preceding sections were taken into account when
designing the MVN instrument.
The main components of the instrument that
should provide an accurate CXB measurement in the
energy range 6-70 keV are:

\begin{enumerate}

\item A periodic aperture blocking system de-
signed to continuously monitor the internal detector
background and realized in the form of a rotating
wheel that blocks the apertures of half the detectors
by the same three-layer material of which the passive
shield of the detector and the collimators themselves
are composed at each instant of time. The wheel
rotation period is 60 s, which should provide insignificant variations in the instrumental background of the
detectors over one aperture modulation period.

\item A 100-$\mu$m-thick beryllium window of the detector that allows the charged particles with energies
in the instrument’s operating range (6-70 кэВ) to be
cut off.

\item  A system of calibration sources for each
detector. The calibration sources ($^{241}$Am) can
be inserted into the collimator on the instrument’s
command to systematically monitor the instrument’s
characteristics, its energy scale, and efficiency. Dur-
ing the planned scientific observations, the calibration
sources are removed from the detector’s field of view
behind the collimator opaque to their photons.

\item A detector thermal stabilization unit. Its
objective is to provide a constant detector temperature, which is very difficult to achieve for an instrument mounted at the external working place of
the ISS due to the large heat flux differences on
its sunlit and unlit sides. The system consists of
heat pipes, heaters, external radiators, and electrically
cooled modules. The operating temperature of the
detectors is $-30^\circ$C. The thermal stabilization unit is
designed so as to provide a constant temperature of
the detector crystals within 0.5 deg, which is very
important for maintaining the instrument’s energy
scale in a maximally stable state $\delta E/E < 0.5$\%.

\end{enumerate}

The MVN instrument is scheduled to operate on-
board the ISS for at least three years. The expected
fraction of the useful time for observations is ∼ 80%;
for the remaining time, the instrument will spend in
the region of the South Magnetic Anomaly and high
geomagnetic latitudes, in which a very high back-
ground count rate of charged particles in the detector
is expected.

At such a fraction of the useful time for
observations and for an expected background count
rate of charged particles $5\times 10^{-3}-5\times10^{-2}$ counts s$^{-1}$ cm$^{-2}$ keV$^{-1}$ , the CXB signal will
be recorded with a significance of 4-6$\sigma$ in a day, 35-
55$\sigma$ in 72 days (the ISS orbital precession period),
and 80-120$\sigma$ in a year.

The CXB detection significance will decrease
with increasing energy due to the falling flux of
CXB photons at a nearly constant (in units of
count s${-1}$ cm$^{-2}$ keV$^{-1}$) background count rate of
events related to the passage of charged particles.
A more detailed prediction of the expected CXB
detection significance over the entire planned onboard
operation of the experiment is presented in Fig. 17.

Over three years of observations, MVN must obtain the CXB surface brightness spectrum in a wide
energy range, 6-70 keV, with a record accuracy that,
together with an accurate calibration of the detector
energy scale and absolute efficiency and the collecting
solid angle of the instrument, will specify a reference
background surface brightness of the Universe for all
observatories in this energy range.
 
\begin{figure}[htb]
\includegraphics[width=\columnwidth,bb=1 163 515 620,clip]{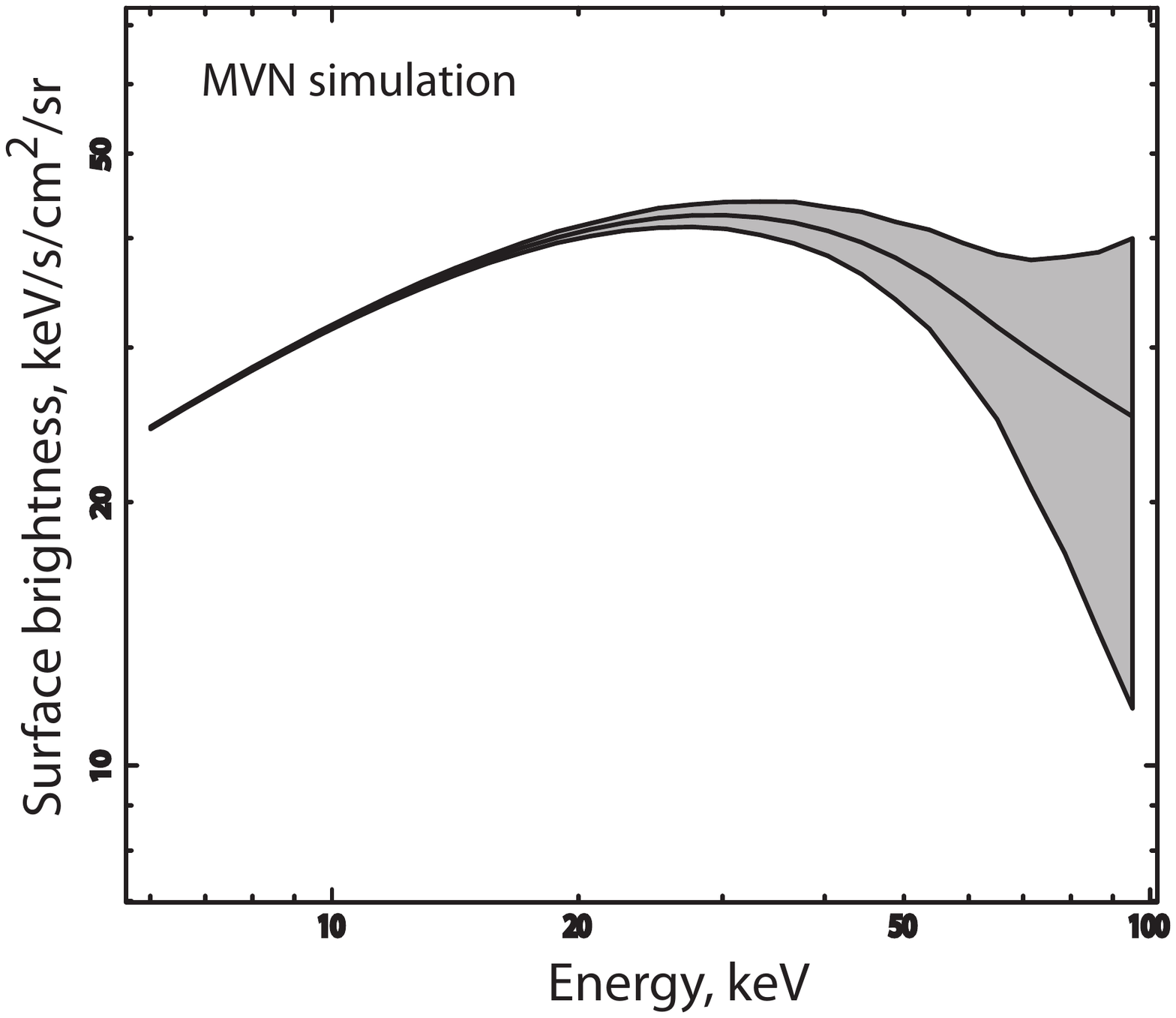}
\caption{\small CXB surface brightness spectrum expected from the operation of the MVN experiment over three years of observations. The gray region indicates the range of expected uncertainties in the surface brightness.}
\label{mvn_spec}
\end{figure}

\section{8. CONCLUSION}

The cosmic X-ray background is one of the fundamental phenomena in X-ray astronomy. The fact
that the CXB emission is the sum of the contributions
from a large number of discrete sources, mostly the
accreting supermassive black holes in the centers of
galaxies at various distances from us (with a minor
contribution of the emission from ordinary galaxies
and galaxy clusters, especially at energies below 2-5 keV), is currently believed to have been firmly established. This means that the CXB emission actually contains information about the history of accretion onto supermassive black holes over the entire
lifetime of the visible Universe. Comparison of the
spectral shape of the CXB surface brightness in the
energy range 1-100 keV with the spectra of accreting supermassive black holes in the nearby Universe
showed that the AGNs observed through the dust
layer obscuring their emission in the standard X-ray
($<$10 keV), ultraviolet, and optical spectral ranges
make a significant contribution to the CXB. Consequently, an accurate measurement of the CXB properties at energies above 2-10 keV makes it possible
to correctly estimate the total mass accumulated in
the black holes at the centers of galaxies.

The flux of X-ray photons from the sky is difficult
to measure, because it is necessary to distinguish
the events in the detectors related to the passage of
X-ray photons from the events related to the passage of charged particles and their derivatives. The
cosmic X-ray background is highly isotropic, which
greatly complicates its reliable separation from the
background of charged particles. The history of CXB
measurements began in the 1960s, with the first experiments in X-ray astronomy. Different techniques
and methodologies, most of which are presented in
this paper, were used at different times to measure
the CXB surface brightness in the energy range 1-100 keV.

The development of an experimental technique in
the X-ray energy range and the advent of semiconductor detectors based on CdTe crystals have allowed
an experiment whose objective would be the most accurate measurement of the CXB surface brightness in
a wide energy range, 6-70 keV, to be proposed. Such
an experiment is currently being developed at the
Space Research Institute of the Russian Academy of
Sciences and being prepared for its installation on the
Russian segment of the International Space Station
in the immediate future.

\bigskip

{\small Author thanks M.N. Pavlinsky, A.A. Lutovinov, and
N.P. Semena for their help in preparing the paper.  The work was supported by RNF grant no. 14-12-01287}

\bibliographystyle{apj}      
\bibliography{xrbreview} 

\hfill  {\sl Translated by V. Astakhov for Astronomy Letters}
\end{document}